\documentclass{IEEEtran}
\usepackage[usenames,dvipsnames,svgnames,table]{xcolor}
\usepackage{pst-plot}
\usepackage{pstricks-add}

\usepackage{amsmath}
\usepackage{amssymb}
\usepackage{amsfonts}
\usepackage{blindtext, graphicx}
\graphicspath{figures}

\usepackage[binary-units=true]{siunitx}
\usepackage[nomessages]{fp}
\usepackage{verbatim}
\usepackage[super]{nth}
\usepackage{soul}
\usepackage{pdf14}
\usepackage{subfig}
\usepackage{multirow}
\usepackage{algpseudocode}
\usepackage{booktabs}

\definecolor{red}{rgb}{1,0,0}
\usepackage{todonotes}

\newcommand{\newred}[1]{}

\newcommand{\keccak}[1]{\textsc{keccak}-$f$[400]\xspace}

\usepackage{mathtools}

\usepackage[dvipsnames]{xcolor}
\usepackage{pifont}%

\usepackage{footnote}

\graphicspath{{./}{./figures/}}

\ifCLASSINFOpdf
\else
\fi
\usepackage{array}
\usepackage{tabulary}
\usepackage{threeparttable}
\usepackage{booktabs}
\usepackage{colortbl}
\usepackage[binary-units=true]{siunitx}
\usepackage{bbm}
\usepackage{subfiles}
\usepackage{tabulary}

\usepackage{changes}
\newcommand{\correction}[1]{#1}
\newcommand{\crossout}[1]{}

\newcommand*\rot{\rotatebox{90}}

\let\OLDthebibliography\thebibliography
\renewcommand\thebibliography[1]{
  \OLDthebibliography{#1}
  \setlength{\parskip}{0.3pt}
  \setlength{\itemsep}{0.3pt plus 0.3ex}
}

\begin{document}
\title{An IoT Endpoint System-on-Chip for Secure\\and Energy-Efficient Near-Sensor Analytics}

\author{
	\IEEEauthorblockN{
		Francesco~Conti,
		Robert~Schilling,
		Pasquale~D.~Schiavone,
		Antonio~Pullini, 
		Davide~Rossi,
		Frank~K.~G\"urkaynak,
		Michael~Muehlberghuber,
		Michael~Gautschi,
		Igor~Loi,
		Germain~Haugou,
		Stefan~Mangard,
		Luca~Benini
	}
    \thanks{
    Francesco~Conti, Pasquale~Schiavone, Antonio~Pullini, Frank~G\"urkaynak, Michael~Muehlberghuber, Michael~Gautschi, Germain~Haugou and Luca~Benini are with ETH~Zurich, Switzerland. Robert~Schilling and Stefan Mangard are with the Graz University of Technology, Austria. Davide~Rossi, Igor~Loi, Francesco~Conti and Luca~Benini are also with the University of Bologna, Italy. Robert~Schilling is also with the Know-Center, Graz, Austria.
    }
    \thanks{
    This work was supported by EU projects MULTITHERMAN (FP7-ERC-291125), ExaNoDe (H2020-671578), OPRECOMP (H2020-732631), and SOPHIA (H2020-ERC-681402).
    }
}

\maketitle

\begin{abstract}
Near-sensor data analytics is a promising direction for IoT endpoints, as it minimizes energy spent on communication and reduces network load
\crossout{
However, performing feature extraction or classification directly on the end-nodes }
\correction{- but it also} poses security concerns, as valuable data \crossout{, “distilled” with application knowledge,} is stored or sent over the network at various stages of the analytics pipeline.
Using encryption to protect sensitive data at the boundary of the on-chip analytics engine is a way to address data security issues.
To cope with the combined workload of analytics and encryption in a tight power envelope, we propose \textit{Fulmine}, a System-on-Chip based on a tightly-coupled multi-core cluster augmented with specialized blocks for compute-intensive data processing and encryption functions, supporting software programmability 
for regular computing tasks.
The \textit{Fulmine} SoC, fabricated in 65\,nm technology, consumes less than 20\,mW on average at 0.8\,V achieving an efficiency of up to 70\,pJ/B in encryption, 50\,pJ/px in convolution, or up to 25\,MIPS/mW in software.
As a strong argument for real-life flexible application of our platform, we show experimental results for three secure analytics use cases: secure autonomous aerial surveillance with a state-of-the-art deep CNN consuming 3.16\,pJ per equivalent RISC op; local CNN-based face detection with secured remote recognition in 5.74\,pJ/op; and seizure detection with encrypted data collection from EEG within 12.7\,pJ/op.
\end{abstract}

\bibliographystyle{IEEEtran}

\vspace{-0.3cm}
\section{Introduction} \label{sec:intro}

The key driver for the development of the Internet-of-Things (IoT) is collecting rich and diverse information streams from sensors, which can then be fed to state-of-the-art learning-based data analytics algorithms.
The information distilled by data analytics on such a rich input set can be used in a virtually unlimited set of applications, such as healthcare or home automation, which have the possibility to change the life of any person for the better \cite{BonomiMCC2012}.
However, in practice, the possibility to seamlessly tap into this rich stream of data is limited by two equally important factors.
First, the amount of data an IoT end-node can extract from sensors and send over the network for analytics is essentially defined by the energy necessary for data transfer itself.
Since IoT end-nodes must work within a tiny power envelope, this fact introduces a significant limit on the volume of data that can be transferred, e.g. the size of captured images, therefore curtailing their usefulness.
Second, due to the ubiquitous nature of IoT devices, they often deal with private or safety critical input data even beyond the predictions of their designers; not only devices such as healthcare wearables acquire potentially safety-critical data, but also seemingly innocuous devices (such as cameras) can potentially acquire highly sensitive information \cite{KumageCC2016}.
To ensure practicality of IoT-based applications, it is imperative that data transmission from end-nodes to the network is protected from data theft or malicious tampering.

To address the first limiting factor, near-sensor smart data analytics is a promising direction; IoT end-nodes must evolve from simple data collectors and brokers into analytics devices, able to perform a pre-selection of potentially interesting data and/or to transform it into a more abstract, higher \textit{information density} form such as a classification tag.
With the burden of \textit{sensemaking} partially shifted from centralized servers to distributed end-nodes, the energy spent on communication and the network load can be minimized effectively and more information can be extracted, making the IoT truly scalable.
However, performing analytics such as feature extraction or classification directly on end-nodes does not address the security concerns. It worsens them: \emph{distilled} data that is stored or sent over the network at several stages of the analytics pipeline is even more privacy-sensitive than the raw data stream~\cite{khan_future_2012,zhang_iot_2014}.
Protecting sensitive data at the boundary of the on-chip analytics engine is a way to address these security issues; however, cryptographic algorithms come with a significant workload, which can easily be of 100-1000s of processor instructions per encrypted byte~\cite{dinu_triathlon_2015}.

This security workload is added to the computational effort imposed by leading feature extraction and classification algorithms, such as deep Convolutional Neural Networks (CNNs).
CNNs are extremely powerful in terms of data analytics, and state-of-the-art results in fields such as computer vision (e.g. object detection~\cite{krizhevsky_imagenet_2012}, scene parsing~\cite{cavigelli_accelerating_2015}, and semantic segmentation tasks~\cite{girshick_rich_2014}) and audio signal analytics~\cite{dahl_context-dependent_2012} have been demonstrated.
While effective, deep CNNs usually necessitate many billions of multiply-accumulate operations, as well as storage of millions of bytes of pre-trained \emph{weights}~\cite{he_deep_2015}. %
The combined workload necessary to tackle these two limitations to the development of smarter IoT - namely, the necessity for \textit{near-sensor analytics} and that for \textit{security} - is formidable, especially under the limited available power envelope and the tight memory and computational constraints of deeply embedded devices.
One possible solution is to augment IoT end-nodes with specialized blocks for compute-intensive data processing and encryption functions while retaining full software programmability to cope with lower computational-intensity tasks.
Specialized processing engines should be tightly integrated both with the software-programmable cores and with one another, streamlining the process of data exchange between the different actors as much as possible to minimize the time and energy spent in data exchange; at the same time, to simplify their usage from the developer's perspective, it should be possible to abstract them, integrating them in standard programming models used in software development for IoT-aware platforms.

In this work, we propose the \SI{65}{\nano\meter} \textit{Fulmine} \textit{secure data analytics} System-on-Chip (SoC), which tackles the two main limiting factors of IoT end-nodes while providing  full programmability, low-effort data exchange between processing engines, (sufficiently) high speed, and low energy.
The SoC is based on the architectural paradigm of tightly-coupled heterogeneous shared-memory clusters~\cite{conti_he-p2012_2014}, where several engines (which can be either programmable cores or specialized hardware accelerators) share the same first-level scratchpad via a low-latency interconnect.
In \textit{Fulmine}, the engines are four enhanced 32-bit OpenRISC cores, one highly efficient cryptographic engine for AES-128 and \textsc{keccak}-based encryption, and one multi-precision convolution engine specialized for CNN computations.
Due to their memory sharing mechanism, cores and accelerators can exchange data in a flexible and efficient way, removing the need for continuous copies between cores and accelerators.
The proposed SoC performs computationally intensive data analytics workloads with no compromise in terms of security and privacy, thanks to the embedded encryption engine.
At the same time, \textit{Fulmine} executes full complex pipelines including CNN-based analytics, encryption, and other arbitrary tasks executed on the processors.

This claim is exemplified in three practical use cases: secure autonomous aerial surveillance in a nano-Unmanned Aerial Vehicle (nano-UAV) consuming \SI{3.16}{\pico\joule} per equivalent RISC operation; on-device CNN-based face detection (as part of a recognition pipeline) with \SI{5.74}{\pico\joule} per operation, including image encryption for external face recognition; and seizure detection with secure data collection within \SI{12.7}{\pico\joule} per operation.
We show that on a workload consisting of balanced contributions from CNNs, AES, and other SW-implementable filters, \textit{Fulmine} provides the best result in terms of \SI{}{\pico\joule}-per-equivalent-RISC-operation, with the nearest state-of-the-art platform in terms of efficiency needing 89$\times$ more time to execute the workload.

The rest of this paper is organized as follows:
\correction{in Section~\ref{sec:related} we compare 
\textit{Fulmine} with the state-of-the-art in low-power IoT computing devices.
Section~\ref{sec:archi} describes the architecture of the SoC; cluster-coupled HW coprocessors are detailed in Section~\ref{sec:accelerators}.
Section~\ref{sec:results} evaluates the implementation results and overall performance, while Section~\ref{sec:results_usecases} focuses on real-world use cases.}
Section~\ref{sec:conclusion} concludes the paper.

\section{State-of-the-Art and Related Work}\label{sec:related}

\crossout{
In this section, we discuss works related to our contribution, i.e. those proposing low-power hardware IPs for encryption or CNNs, and IoT end-node chips that constitute our direct point of comparison.
}
\crossout{Table~\ref{tab:comparison} summarizes the positioning of \textit{Fulmine} with respect to the state-of-the-art, restricted to platforms for which silicon measurements have been published.
}

\subsection{Low-Power Encryption Hardware IPs}
Authenticated encryption is a hot topic in the cryptographic community since it adds additional services on top of data confidentiality.
AES in the Galois Counter Mode~\cite{mcgrew2004security}~(AES-GCM) is one of the most used authenticated encryption schemes today.
For example, Intel added a dedicated finite field multiplication to the AES-NI extension, with a throughput up to 1.03\,cpb~\cite{gueron2013aes}.
However, solutions of this kind are clearly targeting a different scenario from small, low-power IoT devices.

Only a few IoT-oriented commercial AES controllers are available; an example is the Maxim MAXQ1061~\cite{maxim_integrated_maxq1061_2016}, claiming up to 20\,Mbit/s (power consumption data is not currently disclosed).
Research AES accelerators in the sub-100\,mW range for the IoT domain have been proposed by Mathew et al.~\cite{mathew_340_2015} in Intel 22nm technology, Zhang et al.~\cite{zhang_compact_2016} in TSMC 40\,nm and Zhao et al.~\cite{zhao_novel_2015} in 65\,nm; the latter reaches efficiency up to 620 Gbit/s/W thanks to efficient body biasing and a statistical design flow targeted at reducing worst-case guard bands.
A device consuming as little as \SI{0.25}{\micro\watt} for passive RFID encryption has been proposed by Hocquet et al.~\cite{hocquet_harvesting_2011}.
The main differentiating point between our contribution and these hardware encryption techniques is the tightly coupled integration within a bigger low-power system.
\subsection{Low-Power CNN Hardware IPs}
The most common way to accelerate CNNs is to rely on powerful GP-GPUs \cite{cavigelli_accelerating_2015}\cite{ChintalaBenchm}
\correction{ or on FPGAs~\cite{zhang_optimizing_2015}\cite{meloni_high-efficiency_2016}}.
\crossout{
Although able to reach extremely high throughput, this approach is clearly out of the mission profile for IoT end-nodes, due to their power footprint in the order of hundreds of Watts.
}
Some programmable embedded platforms such as ODROID-XU \cite{conti_braininspired_2014}, or Movidius Myriad 2 \cite{Myriad2} improve the energy efficiency of software CNN implementations to up to 120\,Gop/s/W within a few Watts of power envelope, targeting embedded systems such as smartphones or UAVs as well as the booming autonomous car business \cite{Mobileye}.
\correction{
While these platforms are typically powerful enough for embedded scenarios that are not significantly power-constrained (e.g. deep-learning driven autonomous driving), we do not consider them directly comparable to our proposal, since they cannot be used in low-power endnodes: their efficiency is relatively low (up to tens of GMAC/s/W for most GPU and FPGA implementations) and their peak power envelope is typically too high, up to $\sim$\SI{10}{\watt} - $100\times$ the typical envelope considered for endnodes.
}
To the best of our knowledge, the only \correction{deep neural network} commercial solution specifically designed for IoT end-nodes is WiseEye, to be presented by CEVA at CES 2017 \cite{ceva}.

Most research architectures for acceleration of CNNs have focused on specialized architectures to accelerate convolutional layers (e.g. Origami \cite{cavigelli_803_2015}), or convolutional and pooling layers (e.g. ShiDianNao \cite{ShiDianNao} and Eyeriss \cite{EyerissChen2016}).
These accelerators reach efficiencies in the order of a few hundreds of equivalent Gop/s/W.
However, they all rely on highly specialized architectures, their flexibility is limited, and most of them are not capable of implementing the other functionality required by IoT end-nodes, including security and general-purpose signal processing tasks.

One big differentiating point between these platforms are their assumptions in terms of algorithmic and arithmetic accuracy.
Jaehyeong et al. \cite{JaehyeongISSCC2016} rely on 24bit fixed-point arithmetics, but they approximate weights using a low-dimensional representation based on PCA.
Most other works use either 16 bits~\cite{conti_ultralowenergy_2015,park_1.93tops/w_2016} or 12 bits~\cite{cavigelli_803_2015}.
However, recent algorithmic developments such as BinaryConnect~\cite{courbariaux_binaryconnect:_2015} suggest that it is possible to reduce CNN weight precision down to a single bit with limited accuracy losses.
This has been exploited in platforms such as YodaNN \cite{andri_yodann:_2016} to reach efficiency in the order of tens of equivalent Top/s/W.
Another promising approach to improve energy efficiency in classification tasks are extreme learning machines (ELM), based on single-hidden layer feedforward neural networks.
Although they have been proven to consume as little as 0.47 pJ/MAC~\cite{ChenTBCS2016}\cite{YaoTVLSI2016}, their applicability to real-life applications is still restricted to very simple problems.

In this work a flexible approach has been adopted, where the precision of images is fixed to 16 bits, while that of weights can be scaled from 16 to 4 bits.
This approach avoids the requirement of specific training for binary connected networks, while weight precision can be scaled according to the specific application requirements in terms of accuracy, throughput and energy efficiency.

\subsection{End-Node Architectures}
Traditional end-node architectures for the IoT leverage tiny microprocessors, often Cortex-M0 class, to deal with the extreme low-power consumption requirements of applications.
Several commercial solutions have been proposed, among the others, by TI~\cite{_texas-1}, STMicroelectronics~\cite{_stmicroelectronics-1}, NXP~\cite{NXPDualCore}, and Ambiq~\cite{AmbiqApollo}, leveraging aggressive duty-cycling and sub-\SI{10}{\micro\watt} deep-sleep modes to provide extremely low power consumption on average.
Other recent research platforms optimize also the active state, exploiting near-threshold or sub-threshold operation to improve energy efficiency and reduce power consumption during computation~\cite{MyersISSCC2015}\cite{bol_sleepwalker_2013}\cite{IntelVLSI2016}\cite{RoyISQED2016}.

Some commercial architectures leverage lightweight SW acceleration and optimized DSP libraries to improve performance.
The NXP LPC54100~\cite{NXPDualCore} is a commercial platform where a \emph{big} Cortex-M4F core acts as an accelerator for a \emph{little} ultra-low-power Cortex-M0 targeted at always-on applications.
From a software viewpoint, some optimized libraries have been developed to efficiently implement crypto algorithms on Cortex-M3 and M4 architectures, given the criticality of this task for IoT applications.
Examples of these libraries are  SharkSSL~\cite{SharkSSL} and FELICS~\cite{dinu_triathlon_2015}, able to encrypt one block of AES-128-ECB in 1066 cycles and 1816 cycles respectively, both targeting a Cortex-M3.
On the other hand, CMSIS~\cite{CMSIS} is a well-known set of libraries to optimize DSP performance on Cortex-M architectures. 

However, even with software-optimized libraries, these tiny micro-controllers are unfortunately not suitable for secure near-sensor analytics applications using state-of-the-art techniques, which typically involve workloads in the orders of billions of operations per second.
For this reason, a few recent SoCs couple programmable processors with hardwired accelerators, to improve execution speed and energy efficiency in cryptography and other performance-critical tasks. %
In the field of embedded vison, heterogeneous SoCs of this kind include the one recently proposed by Renesas~\cite{RenesasCoolChips2016}, coupling a general purpose processor with an FPU, a DSP, and a signal processing accelerator. Intel~\cite{IntelHOTCHIPS2015} proposed a 14\,nm SoC where a small core with light signal processing acceleration cooperates with a vision processing engine for CNN-based feature extraction and a light encryption engine, within a \SI{22}{\milli\watt} power budget.
Pullini et al. proposed Mia Wallace, a heterogeneous SoC \cite{pullini_heterogeneous_2016} coupling four general purpose processors with a convolutional accelerator.
In the field of bio-signals processing, Konijnenburg et al.~\cite{konijnenburg_multi_2016} proposed a multichannel acquisition system for biosensors, integrating a Cortex-M0 processor and accelerators for digital filtering, sample rate conversion, and sensor timestamping. 
Lee et al.~\cite{LeeJSSC2013} presented a custom bio-signals processor that integrates configurable accelerators for discriminative machine-learning functions (i.e. SVM and active learning) improving energy by up to 145x over execution on CPU.

Similarly to the presented designs, \textit{Fulmine} is a low-power, heterogeneous MPSoC.
In contrast to the other architectures presented here, it tackles at the architectural level the challenge of efficient and secure data analytics for IoT end-nodes, while also providing full programmability with sufficient high performance and low power to sustain the requirements of several near-sensor processing applications.

\section{SoC Architecture}
\label{sec:archi}

\begin{figure}
    \centering
    \includegraphics[clip, trim= 0pt 0pt 0pt 0pt, width=0.42\textwidth]{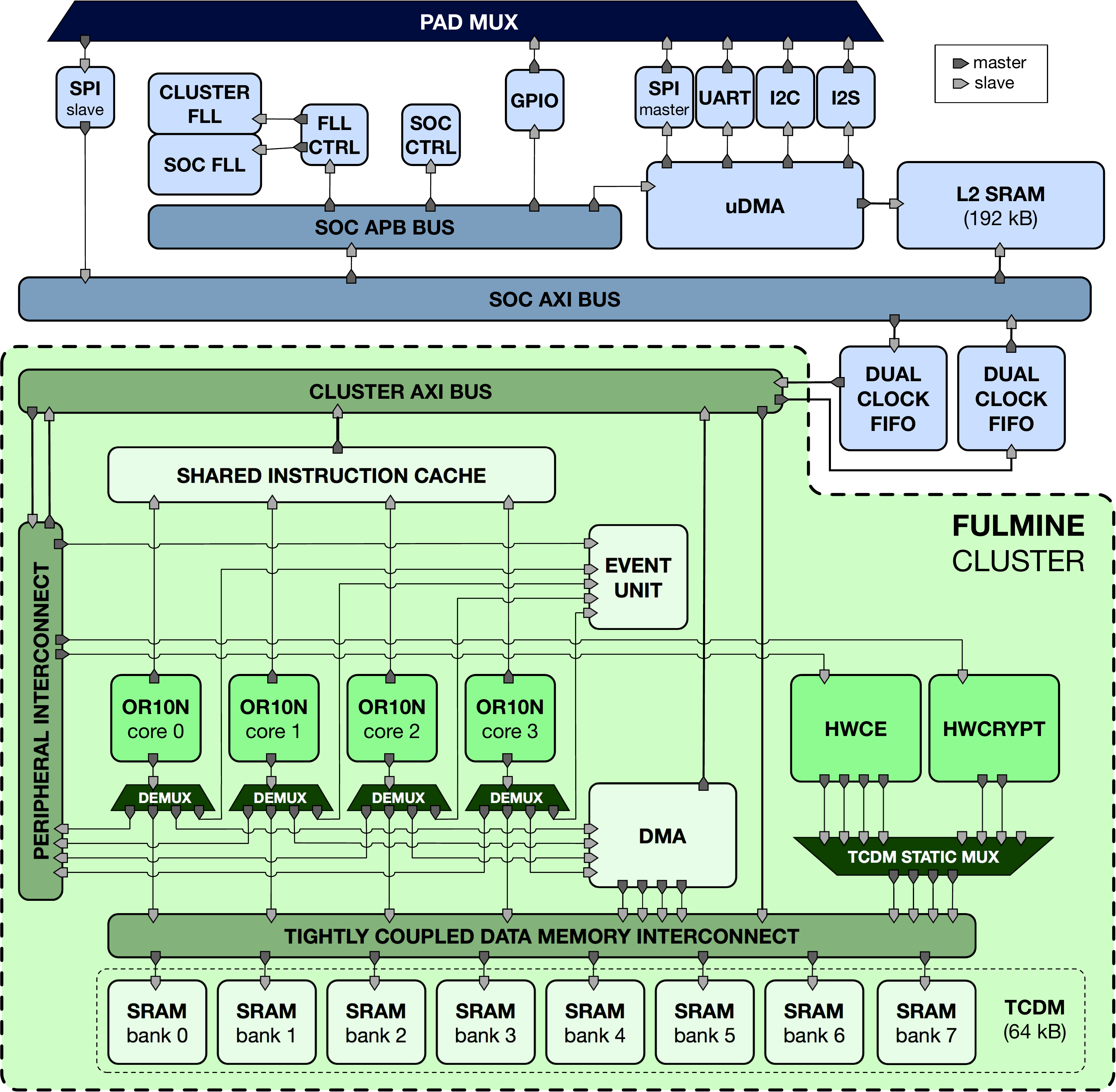}
    \caption{\textit{Fulmine} SoC architecture. The \textsc{soc} domain is shown in shades of blue, the \textsc{cluster} domain in shades of green.}
    \label{fig:fulmine_soc_better}
\end{figure}

The \textit{Fulmine} multi-core System-on-Chip (Figure~\ref{fig:fulmine_soc_better}) implements a secure near-sensor data analytics architecture, which leverages highly efficient processors for software programmable signal processing and control, flexible hardware acceleration for cryptographic functions, convolutional neural networks, and a highly optimized subsystem implementing power management and efficient communication and synchronization among cluster resources.
The architecture, based on the PULP platform~\cite{rossi_pulp:_2015}, is organized in two distinct voltage and frequency domains, \textsc{cluster} and \textsc{soc}, communicating through an AXI4 interconnect and separated by dual-clock FIFOs and level shifters.
Two frequency-locked loops (FLLs) are used to generate clocks for the \crossout{cluster and the rest of the SoC, while the two domains rely on external voltage regulators for power supply.}
\correction{two domains, which rely on separate external voltage regulators for their supply and can be independently power-gated.}
\correction{
The FLLs work with a \SI{100}{\kilo\hertz} external reference clock and support fast switching between different operating modes (less than 10 reference cycles in the worst case).
}

The \textsc{cluster} domain is built around six \textit{processing elements} (four general-purpose processors and two flexible accelerators) that share 64\,kB of level~1 Tightly-Coupled Data Memory (TCDM), organized in eight word-interleaved SRAM banks.
A low-latency logarithmic interconnect~\cite{rahimi_fullysynthesizable_2011} connects all processing elements to the TCDM memory, enabling fast and efficient communication among the resources of the cluster.
The TCDM interconnect supports single-cycle access from multiple processing elements to the TCDM banks; if two masters attempt to access the same bank in the same clock cycle, one of them is stalled using a starvation-free round-robin arbitration policy.
The two hardware accelerators, \textit{Hardware Cryptography Engine} (HWCRYPT) and  \textit{Hardware Convolution Engine} (HWCE), can directly access the same TCDM used by the cores.
This architecture allows data to be seamlessly exchanged between cores and accelerators, without requiring explicit copies and/or point-to-point connections.
To avoid a dramatic increase in the area of the TCDM interconnect, as well as to keep the maximum power envelope in check, the two accelerators share the same set of four physical ports on the interconnect.
The two accelerators are used in a time-interleaved fashion, allowing one accelerator full access to the TCDM at a time, which is suitable for data analytics applications where computation can be divided into several separate stages.

The four OR10N cores are based on an in-order, single-issue, four stage pipeline, implementing the OpenRISC~\cite{_openrisc_2012} instruction set architecture~(ISA), improved with extensions for higher throughput and energy efficiency in parallel signal processing workloads~\cite{gautschi_near-threshold_2016}.
GCC 4.9 and LLVM 3.7 toolchains are available for the cores, while OpenMP 3.0 is supported on top of the bare-metal parallel runtime.
The cores share a single instruction cache of 4\,kB of Standard Cell Memory~(SCM)~\cite{teman_power_2016} that can increase energy efficiency by up to 30\% compared to an SRAM-based private instruction cache on parallel workloads~\cite{rossi_193_2016}.
The ISA extensions of the core include general-purpose enhancements (automatically inferred by the compiler), such as zero-overhead hardware loops and load and store operations embedding pointer arithmetic, and other DSP extensions that can be  explicitly included by means of \textit{intrinsic} calls.
For example, to increase the number of effective operations per cycle, the core includes single instruction multiple data (SIMD) instructions working on 8\,bit and 16\,bit data, which exploit 32\,bit registers as vectors.
Furthermore, the core is enhanced with a native dot-product instruction to accelerate computation-intensive classification and signal-processing algorithms. This single-cycle operation supports both 8\,bit and 16\,bit vectors using two separate datapaths to reduce the timing pressure on the critical path.
Fixed point numbers are often used for embedded analytics and signal processing applications; for this reason, the core has also been extended with single-cycle fixed point instructions including rounded additions, subtractions, multiplications with normalization, and clipping instructions. 

The cluster features a set of peripherals including a direct memory access (DMA) engine, an event unit, and a timer. 
The processors can access the control registers of the hardware accelerators and of the other peripherals through a memory mapped interface implemented as a set of private, per-core demultiplexers (DEMUX), and a peripheral interconnect shared among all cores.
\crossout{
Latency-critical peripherals such as the event unit and the DMA are directly connected to the DEMUXes, thereby guaranteeing very low latency access (i.e. one cycle) and eliminating contention when accessing their control ports.
The other peripherals are mapped on the peripheral interconnect through an additional pipeline stage to remove them from the critical path of the core.
As hardware accelerators can be busy executing jobs for hundreds to thousands of cycles with no or little need for interaction, their control ports are accessed through this lower-priority path.
}
The peripheral interconnect implements the same architecture of the TCDM interconnect, featuring a different addressing scheme to provide 4\,kB of address map for each peripheral.

The DMA controller available in the cluster is an evolution of the one presented in~\cite{rossi_ultralowlatency_2014}, and enables fast and flexible communication between the TCDM and the L2 memory trough four dedicated ports on the TCDM interconnect and an AXI4 plug on the cluster bus.
\crossout{
The DMA has a set of private per-core command FIFOs accessed through the DEMUXes that converge on a larger global command queue through a round robin arbiter.
Hence, all the cores within the cluster can asynchronously and concurrently push transfers to the control ports, avoiding the usage of inefficient software locks to guarantee mutual access to the DMA internal resources.
Moreover, i}
In contrast to traditional memory mapped interfaces, access to the internal DMA programming registers is implemented through a sequence of control words sent to the same address, significantly reducing DMA programming overheads (i.e. less then 10 cycles to initiate a transfer, on average).
The DMA supports up to 16 outstanding 1D or 2D transfers to hide L2 memory latency and allows 256 byte bursts on the 64-bit AXI4 interface to guarantee high bandwidth.
Once a transfer is completed, the DMA generates an event to the cores that can independently synchronize on any of the enqueued transfers by checking the related transfer ID on the DMA control registers.
Synchronization of DMA transfers and hardware accelerated tasks is hardware-assisted by the event unit.
The event unit can also be used to accelerate the typical parallelization patterns of the OpenMP programming model, requiring, for example, only 2 cycles to implement a \textit{barrier}, 8 cycles to open a \textit{critical section}, and 70 cycles to open a \textit{parallel section}.
These features are all essential to guarantee high computational efficiency during execution of complex tasks such as CNNs in \textit{Fulmine}, as detailed in Section \ref{sec:accelerators}.

The \textsc{soc} domain contains 192\,kB of L2 memory for data and instructions, a 4\,kB ROM, a set of peripherals, and a power management unit.
Furthermore, the \textsc{soc} includes a (quad) SPI master, I2C, I2S, UART, GPIOs, a JTAG port for debug, and a (quad) SPI slave that can be used to access all the SoC internal resources.
An I/O DMA subsystem (uDMA) allows to autonomously copy data between the L2 memory and the external interfaces, even when the cluster is in sleep mode.
This mechanism allows us to relieve cores from the frequent control of peripherals necessary in many microcontrollers, and to implement a double buffering mechanism both between IOs and L2 memory and between L2 memory and TCDM.
Therefore, I/O transfers, L2 memory to TCDM transfers, and computation phases can be fully overlapped.

\begin{figure}
    \centering
    \includegraphics[clip, trim= 0pt 0pt 0pt 0pt, width=0.48\textwidth]{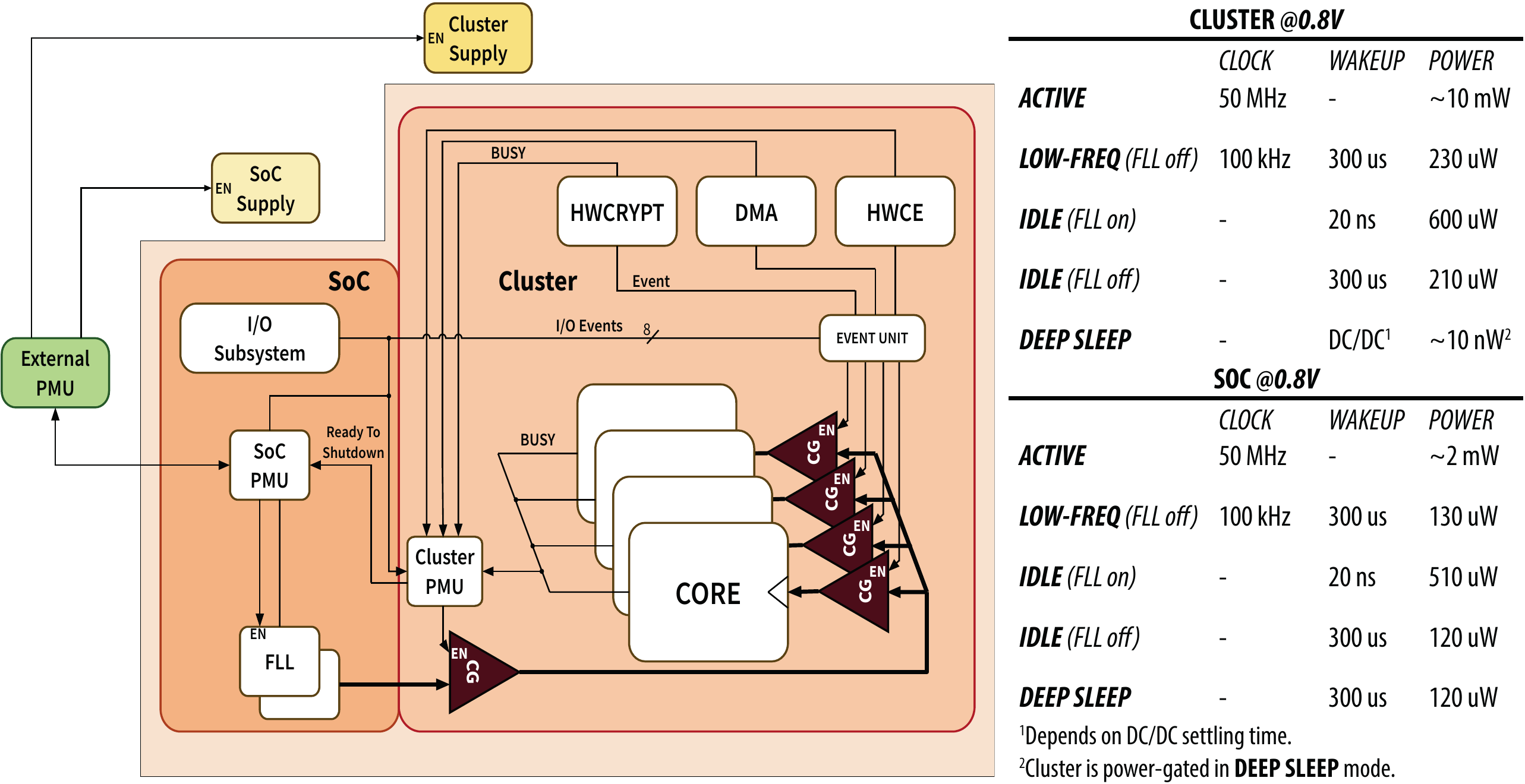}
    \caption{\textit{Fulmine} power management architecture and power modes.}
    \label{fig:power_manager}
\end{figure}

A sophisticated power management architecture distributed between the \textsc{soc} and \textsc{cluster} domains can completely clock-gate all the resources when idle, as shown in Figure~\ref{fig:power_manager} (\textit{idle} mode with FLL on).
\correction{The power manager can also be programmed to put the system in a low power retentive state by switching down the FLLs and relying on the low-frequency reference clock (\textit{low freq} and \textit{idle} mode).
Finally, it can be used to program the external DC/DC converter to fully power-gate the \textsc{cluster} domain.
}
The event unit is responsible for automatically managing the transitions of the cores between the active and idle state.
\crossout{Processors are put in idle by reading a configuration register of the event unit through their private port (mapped on the DEMUX).
This happens when the processors execute an explicit \textit{Wait For Event} instruction, for example during a synchronization barrier or after a DMA transfer.
The processors are then stalled by the event unit, and once all pending transactions (e.g., cache refills) are complete, they are clock-gated.
}
\correction{
To execute a \textit{wait-for-event} instruction, the cores try to read a special register in the event unit; this load is kept stalled until the event comes so that the core pipeline is stalled in a known state.
After pending transactions and cache refills are complete, the event unit gates the core clock.
}
\crossout{A clock gate management unit probes the state of all engines of the cluster: the processors, the hardware accelerators, and the DMA.
If none of the resources are busy, the SoC power manager is notified.
It reacts accordingly to the policy stored in its configuration register: it
}
\correction{The clock gating manager}
gates the cluster clock if the \textit{idle} mode is selected \correction{and no engine is busy}, or it activates the handshaking mechanism with the external regulator to power gate the cluster if the \textit{deep-sleep} mode is selected.
Once the wake-up event reaches the power management unit, the latter reactivates the cluster, then it forwards the event notification to the event unit, waking up the destination core \correction{on turn}.
\crossout{
\textsc{soc} and \textsc{cluster} events can be generated by all SoC peripherals (including GPIOs), by hardware accelerators, by the DMA, by the cluster timer, or by processors through the configuration interface of the event unit mapped in the peripheral interconnect.
}
\crossout{Table \ref{tab:power_modes}}\correction{Figure \ref{fig:power_manager}} reports all the power modes along with their average wakeup time and power consumption, divided between the \textsc{cluster} and \textsc{soc} domains.
\crossout{
A programmable frequency-locked loop (FLL) with an external reference clock of \SI{0.1}{\mega\hertz} is used for each domain of \emph{Fulmine}. The FLL supports a fast frequency switch between different operating modes for the cluster, and in the worst case is able to complete the frequency switch in less than 10 cycles of the reference clock. To perform the switch, the cluster is first put in sleep mode, then woken up again when the FLL locks. As sleep and wakeup can employ the fast mechanism previously described, the frequency switch can be performed in as little as \SI{10}{\micro\second}.
}
\correction{
As the \textsc{cluster} and \textsc{soc} power domains are managed independently, it is possible to transparently put the \textsc{cluster} in \textit{idle}, where it consumes less than \SI{1}{\milli\watt}, when waiting for an event such as the end of an I/O transfer to L2 or an external interrupt that is expected to arrive often.
It is possible to partially trade off wakeup time versus power by deciding whether to keep the FLLs active in \textit{idle} mode: by paying a $\sim$\SI{400}{\micro\watt} cost, wakeup time is reduced to essentially a single clock cycle (\SI{20}{\nano\second}), versus a maximum of 10 reference cycles ($\sim$\SI{320}{\micro\second}) if the FLL is off.
The \textit{deep sleep} mode instead enables efficient duty cycling in the case computing bursts are relatively rare, by completing power-gating the \textsc{cluster} domain and keeping the \textsc{soc} domain in a clock-gated, retentive state.
}

\section{Cluster-coupled accelerator engines}
\label{sec:accelerators}

\correction{
In this Section we describe in detail the architecture of the two cluster-coupled accelerator engines, HWCRYPT and HWCE.
The main purpose of these engines is to provide a performance and efficiency boost on computations, and they were designed to minimize active power, e.g. by using aggressive clock gating on time-multiplexed sub-modules and by making use of latches in place of regular flip-flops to implement most of the internal buffering stages.
}

\crossout{
The shared-memory architecture of the HWCRYPT and HWCE accelerators enables efficient zero-copy data exchange with the cores and the DMA engine, while the event unit can be leveraged to efficiently switch computation from cores to accelerators and vice-versa and/or to trigger data transfers to/from the L2.
These features, together with a highly optimized software runtime, are essential to guarantee a high level of programmability and highly efficient execution of complex tasks such as CNNs, which require complex computation patterns, frequent memory transfers for data set tiling, and synchronization between all the processing elements to implement efficient parallelization.
}
\correction{
The shared-memory nature of the HWCRYPT and HWCE accelerators enables efficient zero-copy data exchange with the cores and the DMA engine, orchestrated by the cluster event unit.
This architecture enables complex computation patterns with frequeny transfers of data set tiles from/to memory.
}
A typical application running on the \textit{Fulmine} SoC operates conceptually in the following way.
First, the input set (e.g. a camera frame) is loaded into the L2 memory from an external I/O interface using the uDMA.
The cluster can be left in sleep mode during this phase and woken up only at its conclusion.
The input set is then divided into tiles of appropriate dimension so that they can fit in the L1 shared TCDM; one tile is loaded into the cluster, where a set of operations are applied to it either by the SW cores or the HW accelerators.
These operations can include en-/decryption and convolutions (in HW), plus any SW-implementable filter.
The output tiles are then stored back to L2 memory using DMA transfers, and computation continues with the next tile.
Operations such as DMA transfers can typically be overlapped with computation by using double buffering to reduce the overall execution time.

\crossout{
Cores are used both for actual computation on the data set and for control; to avoid inefficient busy waiting, events are employed by the HW accelerators to notify completed execution.
Accelerator events trigger an appropriate interrupt in the controller core while it is either in sleep and clock-gated, or executing a filter of its own in parallel to HW-accelerated computation.
For example, a secured convolutional layer of a CNN can comprise full decryption of all inputs, accumulation of convolutions, pooling, non-linear activation and encryption of all outputs, requiring both accelerator control and software-based computation.
}

\subsection{Hardware Encryption Engine}
\label{sec:hwcrypt}

\begin{figure}[b]
    \centering
    \includegraphics[clip, trim= 0pt 0pt 0pt 0pt, width=0.36\textwidth]{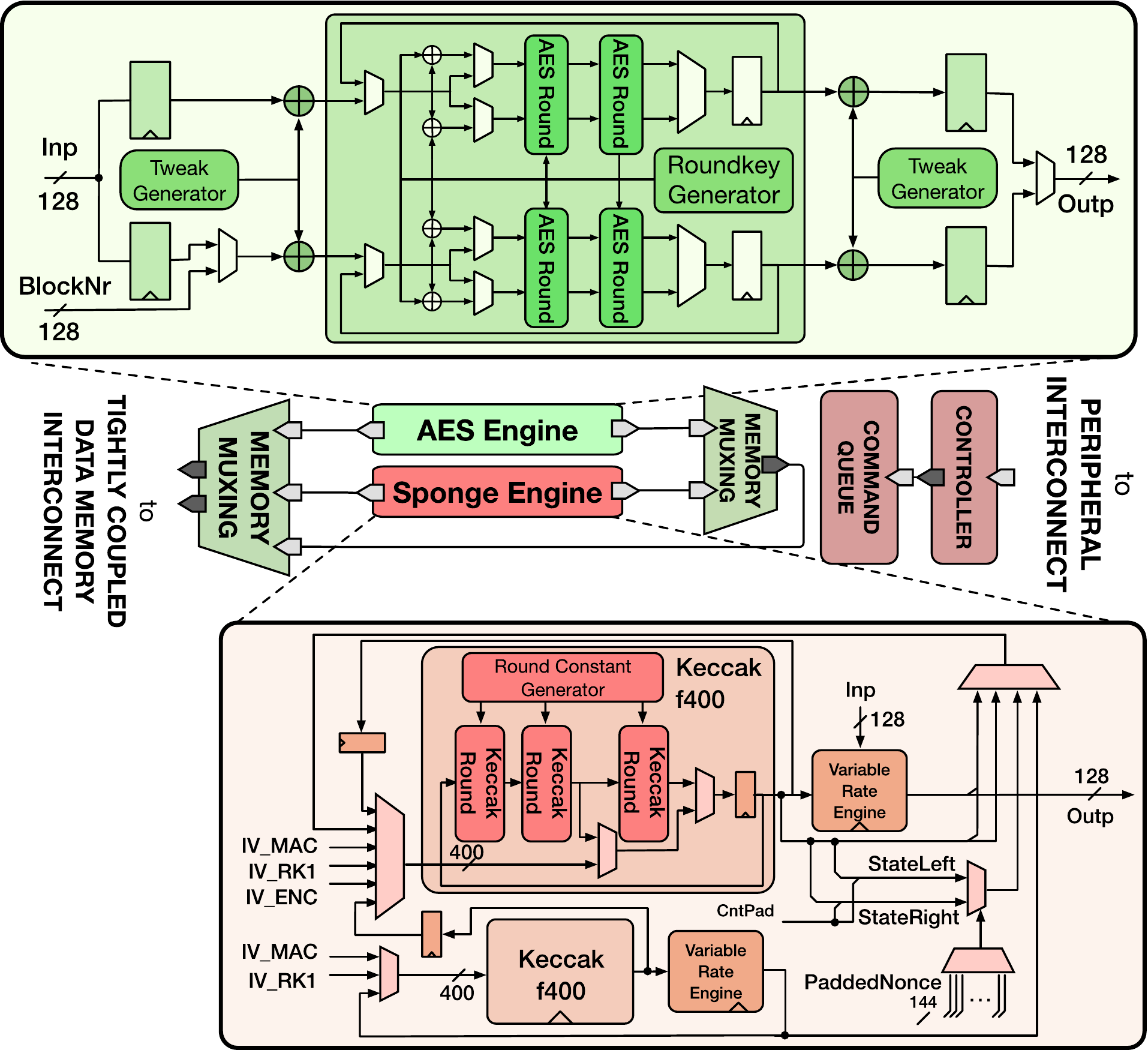}
    \caption{HWCRYPT datapath overview, with details of the AES-128 and the sponge engine.}
    \label{fig:fulmine_hwcrypt_dp}
\end{figure}

The Hardware Encryption Engine~(HWCRYPT), as shown in Figure~\ref{fig:fulmine_hwcrypt_dp}, implements a dedicated acceleration unit for a variety of cryptographic primitive operations, exploiting the advantages of the shared memory architecture of the SoC.
The HWCRYPT is based on two parallel cryptographic engines, one implementing the AES-128~\cite{nist_advanced_2001} block cipher and the other one implementing the \keccak{}~\cite{bertoni_keccak_2009} permutation (a smaller version of the SHA-3 permutation) used in a flexible sponge construction.
The AES-128 engine includes two instances of a round-based AES-128 design with a shared on-the-fly round-key computation module.
Each of the two AES-128 instances is based on two cipher rounds supporting both encryption and decryption.
The round-key generator keeps track of the last round-key during encryption operations, which acts as the starting point to generate round-keys for a decryption operation.
The AES-128 engine of the HWCRYPT implements the Electronic-Code-Book~(ECB) mode as well as the XEX-based tweaked-codebook mode with ciphertext stealing (XTS)\crossout{, which is a recommended mode of operation for disk encryption by the National Institute of Standards and Technology}~\cite{dworkin_recommendation_2010}.
XTS uses two different encryption keys, one to derive the initial tweak and the other one to encrypt the data.
When using the same key for deriving the initial tweak and encrypting the data, the encryption scheme is changed to XEX~\cite{rogaway_efficient_2004} without implications to the overall security.
Furthermore, the accelerator supports the individual execution of a cipher round similar to the Intel AES-NI instructions~\cite{gueron_intel_2010} to boost the software performance of other new AES round-based algorithms~\cite{wu_aegis:_2013,hoang_robust_2015}.

Although AES-128-ECB is a fast encryption mode, it is not recommended to use it to encrypt larger blocks of data.
Since every block is encrypted independently using the same key, the same plaintext always yields the same ciphertext, which reveals patterns of the plaintext in the ciphertext.
To overcome this issue, the AES-128-XTS mode \crossout{(Figure~NA)} uses a so-called tweak $T$ to modify the encryption for each block.
The encryption function $E$ in the block diagram denotes an AES-128-ECB encryption with the key $K_1$ and $K_2$ respectively.
The tweak is XORed to the plaintext and to the resulting ciphertext of the AES-128-ECB encryption.
Since the tweak is different for each block of plaintext, it is denoted as $T_i$ for the i-th block of data.
The initial tweak $T_0$ is computed by encrypting the sector number $SN$, 
derived from the address of the data, using the encryption key $K_1$ and multiplying it with $\alpha^i$ with $i = 0$. The multiplication with $\alpha^i$ ensures that the tweak is different for each block.
\correction{
The XTS mode is defined by Equation~\ref{eq:arch:xts_basic}:
 \begin{align} 
   \label{eq:arch:xts_basic}
   T_i &= E_{K_1}(SN) \otimes \alpha^i\\
   C_i &= E_{K_2}(P_i \oplus T_i) \oplus T_i \nonumber
 \end{align}
}
\crossout{
The depicted XTS mode in Figure~NA is represented in a mathematical form in Equation~\ref{eq:arch:xts_basic}.
}
The address-dependent tweak $T_i$ is derived by a multiplication between the initial tweak and $\alpha^i$.
The multiplication is performed in the finite \correction{Galois} field\footnote{In the following,
$\otimes$ denotes the 128-bit finite field multiplication in which also the exponentiation is performed.} \crossout{or Galois field }GF$\left(2^{128}\right)$ defined by the irreducible polynomial $x^{128} + x^7 + x^2 + x + 1$.
AES-128-XTS requires a 128-bit finite field multiplier and exponentiator, which is rather complex in terms of VLSI implementation.
To reduce this complexity, we first observe that $\alpha$ is constant with the recommended value $\alpha = 2$.
\crossout{
We can derive the tweak for the current block sequentially as a function of the previous tweak $T_{i-1}$, turning the exponentiation into a sequential multiplication by two as shown in Equation~\ref{eq:arch:xts_tweak_iterative}.
}
\correction{
$T_{i}$ is derived from $T_{i-1}$ as a one-bit shift with a conditional XOR with the irreducible polynomial, i.e. $T_i = T_{i-1} \otimes 2$, which implements the multiplication by 2 in the Galois field $GF(2^{128})$.
}
\crossout{
In the finite field GF$\left(2^{128}\right)$, a multiplication by two is much simpler and can be realized with a shift by one to the left with a conditional XOR with the irreducible polynomial.
}

The sponge engine implements two instances of the \keccak{} permutation, each based on three permutation rounds.
\keccak{}'s architecture is optimized to match the length of the critical path of the AES-128 engine.
Permutations support a flexible configuration of the rate and round parameters. 
The rate defines how many bits are processed within one permutation operation, and it can be configured from 1\,bit to 128\,bits in powers of two.
This parameter supports a trade-off between security and throughput.
The more bits are processed in one permutation call, the higher the throughput - but with a cost regarding the security margin of the permutation.
The round parameter configures the number of \keccak{} rounds applied to the internal state.
It can be set up as a multiple of three or for 20 rounds as defined by the specification of \keccak{}.
The two instances of permutations are combined to implement an authenticated encryption scheme based on a sponge construction with a prefix message authentication code that additionally provides integrity and authenticity on top of confidentiality.
\crossout{In Figure~\ref{fig:sponge_enc}, we show the sponge construction for encryption. Initially, the}
\correction{In the sponge construction for encryption, the initial} state of the sponge is filled with the key $K$ and the initial vector $IV$.
After executing the \keccak{} permutation $p$, we sequentially squeeze an encryption pad and apply the permutation function to encrypt all plaintext blocks $P_i$ via an XOR operation. 
Apart from this favorable mode of operation, the sponge engine also provides encryption without authentication and direct access to the permutations to allow the software to accelerate any \keccak{}-based algorithm.

The HWCRYPT utilizes two 32\,bit memory ports of the TCDM interconnect, while an internal interface performs the conversion from 32\,bit to the 128\,bit format used by the encryption engines. The system is designed so that memory interface bandwidth matches the requirements of all cipher engines.
The HWCRYPT is programmed and started through a dedicated set of configuration registers, which allows the reconfiguration of a new encryption operation while the HWCRYPT is busy by using a command queue that supports up to four pending operations. 
The current state of the HWCRYPT can be monitored via status registers.
The accelerator supports a flexible event and interrupt system to indicate when one or all operations have finished.

\subsection{Hardware Convolution Engine}
\label{sec:hwce}

\begin{figure*}
    \centering
    \includegraphics[clip, trim= 0pt 0pt 0pt 0pt, width=0.82\textwidth]{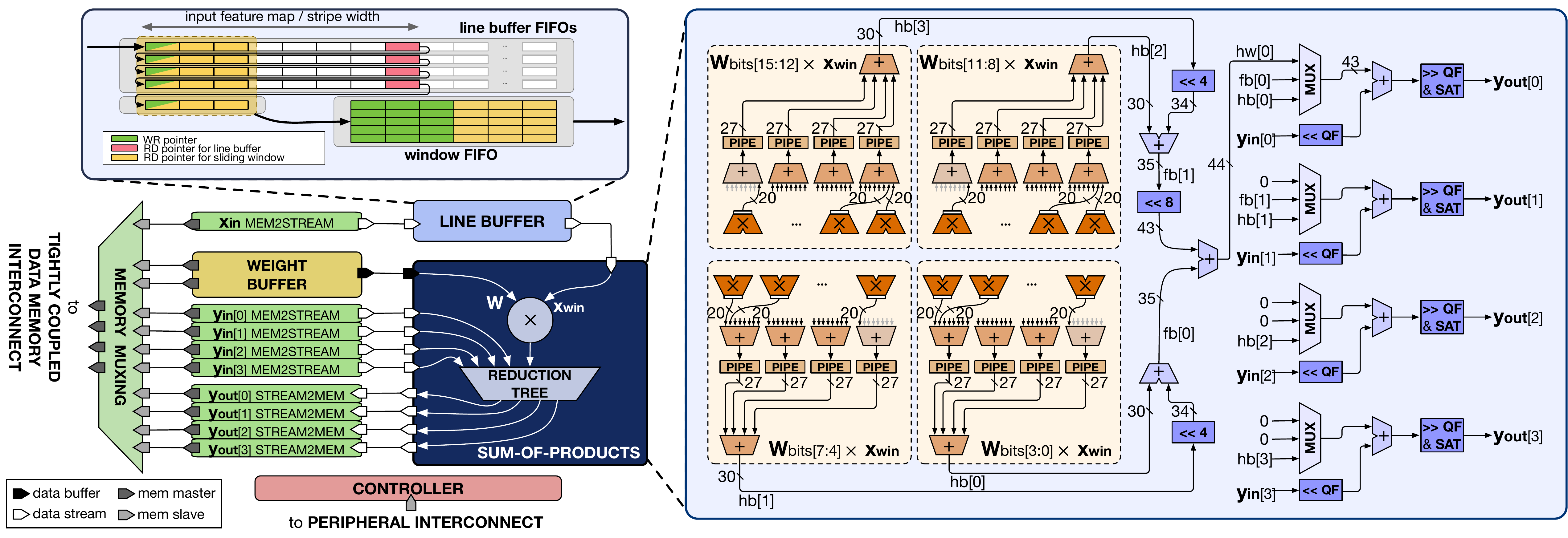}
    \caption{\textit{Fulmine} HWCE architecture, with the \textit{controller} shaded in red, the \textit{wrapper} in green, and the \textit{datapath} in blue. The diagram also shows details of the line buffer and sum-of-products submodules microarchitecture.}
    \label{fig:fulmine_hwce_full_dp}
\end{figure*}

The Hardware Convolution Engine (HWCE) is based on a precision-scalable extension of the design proposed by Conti and Benini~\cite{conti_ultralowenergy_2015} to accelerate convolutional layers in deep CNNs. %
These layers, which constitute the overwhelming majority of computation time in most CNNs~\cite{cavigelli_accelerating_2015}, are composed of a linear transformation that maps $N_\mathrm{if}$ input feature maps into $N_\mathrm{of}$ output feature maps by means of a set of convolutional filters; and a pointwise non-linear activation function, often a rectifier (ReLU) \crossout{or a hyperbolic tangent}.
The linear part of convolutional layers is usually the dominant operation by far; its typical form for ${ k_\mathrm{\scriptscriptstyle of} \in 0\cdots N_\mathrm{\scriptscriptstyle of}-1}$ is the following:
\begin{equation}
	\mathbf{y}({\scriptstyle k_\mathrm{\scriptscriptstyle of}}) =
	\mathbf{b}({\scriptstyle k_\mathrm{\scriptscriptstyle of}})
	+ 
	\sum_{k_\mathrm{\scriptscriptstyle if}=0}^{N_\mathrm{\scriptscriptstyle if}-1}
	\Big(\mathbf{W}({\scriptstyle k_\mathrm{\scriptscriptstyle of}}, {\scriptstyle k_\mathrm{\scriptscriptstyle if}}) * 
	\mathbf{x}({\scriptstyle k_\mathrm{\scriptscriptstyle if}})
	\Big).
	\label{eq:conv_layer}
\end{equation}
The specific task executed by the HWCE is the acceleration of the \textit{accumulation of convolutions} that are at the core of Equation \ref{eq:conv_layer}.
To represent input/output pixels and weights, a fixed-point data representation with 16\,bits is used by default.
The number of fractional bits is configurable at run time.
HWCE can natively perform $5\times 5$ and $3\times 3$ convolutions, and any arbitrary convolution by combining these two in software.
The key novelty in the HWCE design with respect to~\cite{conti_ultralowenergy_2015} is that the \textit{Fulmine} HWCE can exploit the relative insensitivity of CNNs to weight approximation~\cite{courbariaux_training_2014}\cite{courbariaux_binaryconnect:_2015} by reducing the arithmetic precision of the convolution weights to 8 or 4\,bit.
In that case, the internal datapath is reconfigured so that two or four convolutions respectively (on different output $k_\mathrm{\scriptscriptstyle of}$ feature maps) are computed simultaneously, while feature map pixels still use the full 16\,bit representation.
In these scaled precision modes, a similar level of accuracy to the 16\,bit full precision CNNs can be maintained by proper training, with access to significantly improved performance, memory footprint, and energy efficiency as is shown in Section~\ref{sec:results}.

Figure~\ref{fig:fulmine_hwce_full_dp} depicts the HWCE architecture, which can be divided into three main components: a \textit{datapath} performing the main part of the data plane computation in a purely streaming fashion\correction{, relying on an AXIStream-like handshake for back-pressure}; a \textit{wrapper} that connects and decouples the datapath streaming domain from the memory-based cluster; and a \textit{controller} that provides a control interface for the accelerator.
In the full-precision 16\,bit mode, the sum-of-products datapath is used to perform a convolution between a preloaded filter $\mathbf{W}$ (stored in a weight buffer) and a 5$\times$5 $\mathbf{x}_\mathrm{win}$ window extracted from a linear $\mathbf{x}$ input feature map stream.
Window extraction is performed by a line buffer, which is realized with latch-based SCMs for optimized energy efficiency.
The line buffer is composed by a set of FIFO queues with two read pointers: one to implement the mechanism to pass the oldest pixel to the next FIFO and the other to extract the 5$\times$5 sliding window.
The output of the sum-of-products is summed to an input pre-accumulated $\mathbf{y}_\mathrm{in}$ value; in other words, the accelerator needs no internal memory to perform the feature map accumulation component of Equation~\ref{eq:conv_layer} but uses directly the shared memory of the cluster.
The wrapper, shaded in green in Figure~\ref{fig:fulmine_hwce_full_dp} is responsible for generating memory accesses through four memory ports to the TCDM to feed the accelerator $\mathbf{x}$,$\mathbf{y}_\mathrm{in}$ streams and write back $\mathbf{y}_\mathrm{out}$ (partial) results.
The controller (red in Figure~\ref{fig:fulmine_hwce_full_dp}) contains a register file which can host a queue of two jobs, each consisting of pointers to $\mathbf{x}$, $\mathbf{W}$, $\mathbf{y}$, strides for the wrapper address generators, and other configuration such as the number of fractional bits to use.
The controller is mapped in the cluster peripheral interconnect.

To support three different possible sizes for weights, the HWCE sum-of-products datapath must be able to perform 16\,bit $\times$ 16\,bit products as well as 8\,bit $
\times$ 16\,bit and 4\,bit $\times$ 16\,bit ones.
The two or four filters hosted in the weight buffer in scaled precision modes are not consecutive, but they are interleaved: in full precision mode a location represents a single 16\,bit weight; in the scaled precision modes, it represents two 8\,bit or four 4\,bit weights.
The sum-of-products datapath is designed in a hierarchical way to maximize its reuse between the three configurations.
Four submodules (shown in orange in Figure~\ref{fig:fulmine_hwce_full_dp}) compute the sum-of-products of $\mathbf{x}_\mathrm{win}$ with a 4\,bit slice of $\mathbf{W}$ each, using a set of signed multipliers and a first-stage reduction tree.
A second-stage reduction tree and a set of multiplexers are used to combine these four partial sums-of-products to produce one, two or four concurrent $\mathbf{y}_\mathrm{out}$ outputs; fractional part normalization and saturation are also performed at this stage.
As multiple accumulations of convolutions are performed concurrently, the $\mathbf{y}_\mathrm{in}$ and $\mathbf{y}_\mathrm{out}$ streamers are replicated four times.
All HWCE blocks are aggressively clock gated so that each component consumes power only when in active use.

\section{Experimental evaluation}
\label{sec:results}

\begin{figure*}[tb!]
  \centering
  \subfloat[Cluster maximum frequency in three operating modes.\label{fig:cluster_freq}] {
    \centering
    \includegraphics[clip, trim= 0pt 0pt 0pt 0pt, width=0.4\textwidth] {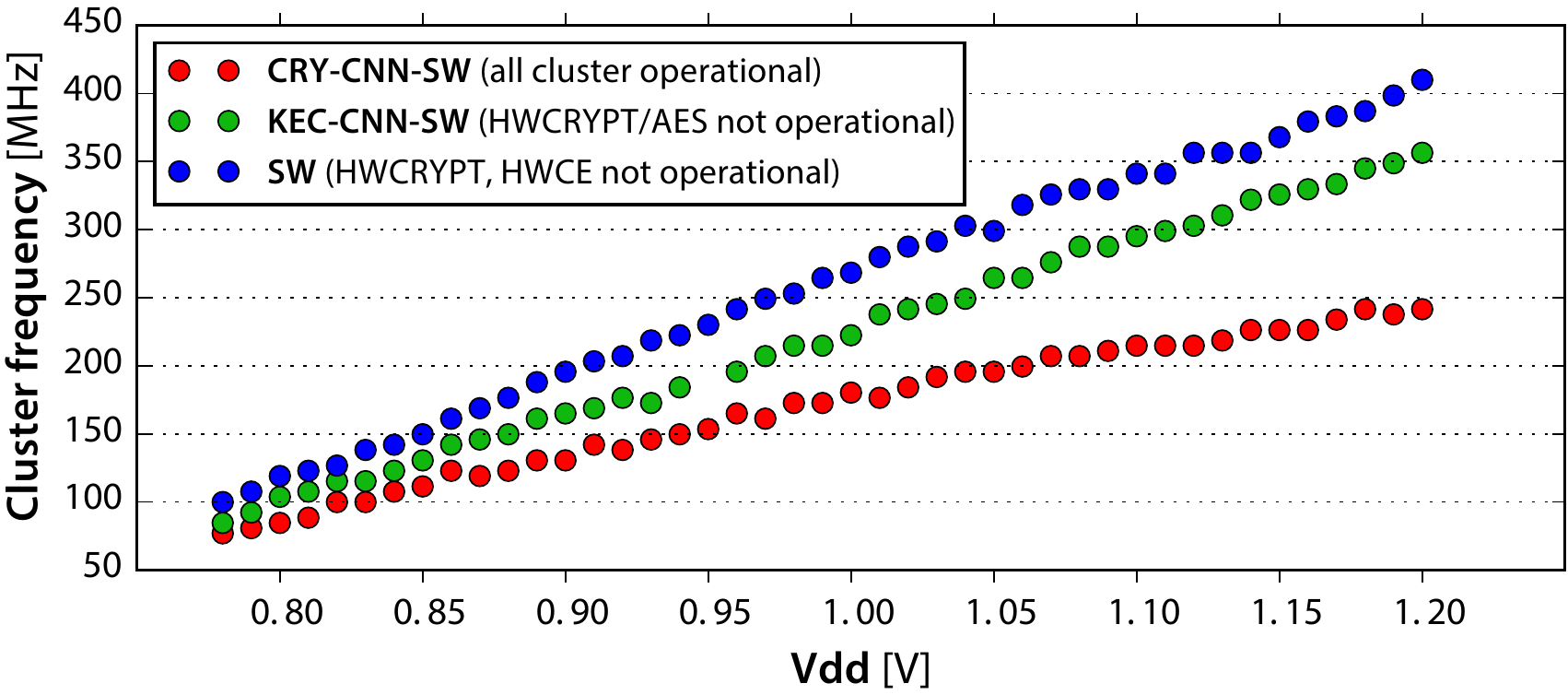}
  }
  \quad
  \subfloat[Cluster power at maximum frequency in three operating modes.\label{fig:cluster_pow}] {
    \centering
    \includegraphics[clip, trim= 0pt 0pt 0pt 0pt, width=0.4\textwidth] {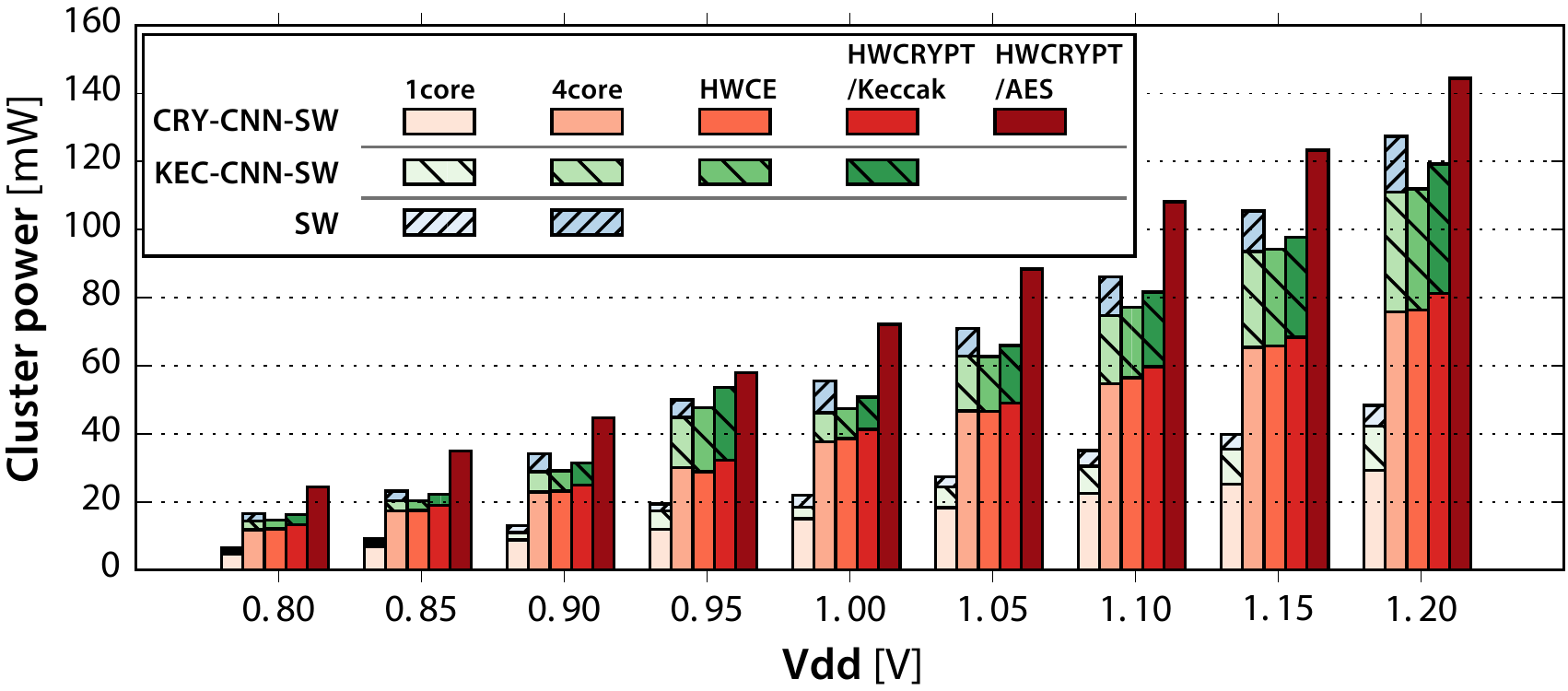}
  }
  \caption{Cluster maximum operating frequency and power in the {\textsc{cry-cnn-sw}}, {\textsc{kec-cnn-sw}}, and {\textsc{sw}} operating modes. Each set of power bars, from left to right, indicates activity in a different subset of the cluster. {\textsc{kec-cnn-sw}} and {\textsc{sw}} bars show the additional power overhead from running at the higher frequency allowed by these modes.}
  \label{fig:cluster_op}
\end{figure*}

\begin{figure}[tb!]
  \centering
  \subfloat[HWCRYPT time/energy per byte.\label{fig:hwcrypt_cpb_eff}] {
    \centering
    \includegraphics[clip, trim= 0pt 0pt 0pt 0pt, width=0.22\textwidth]{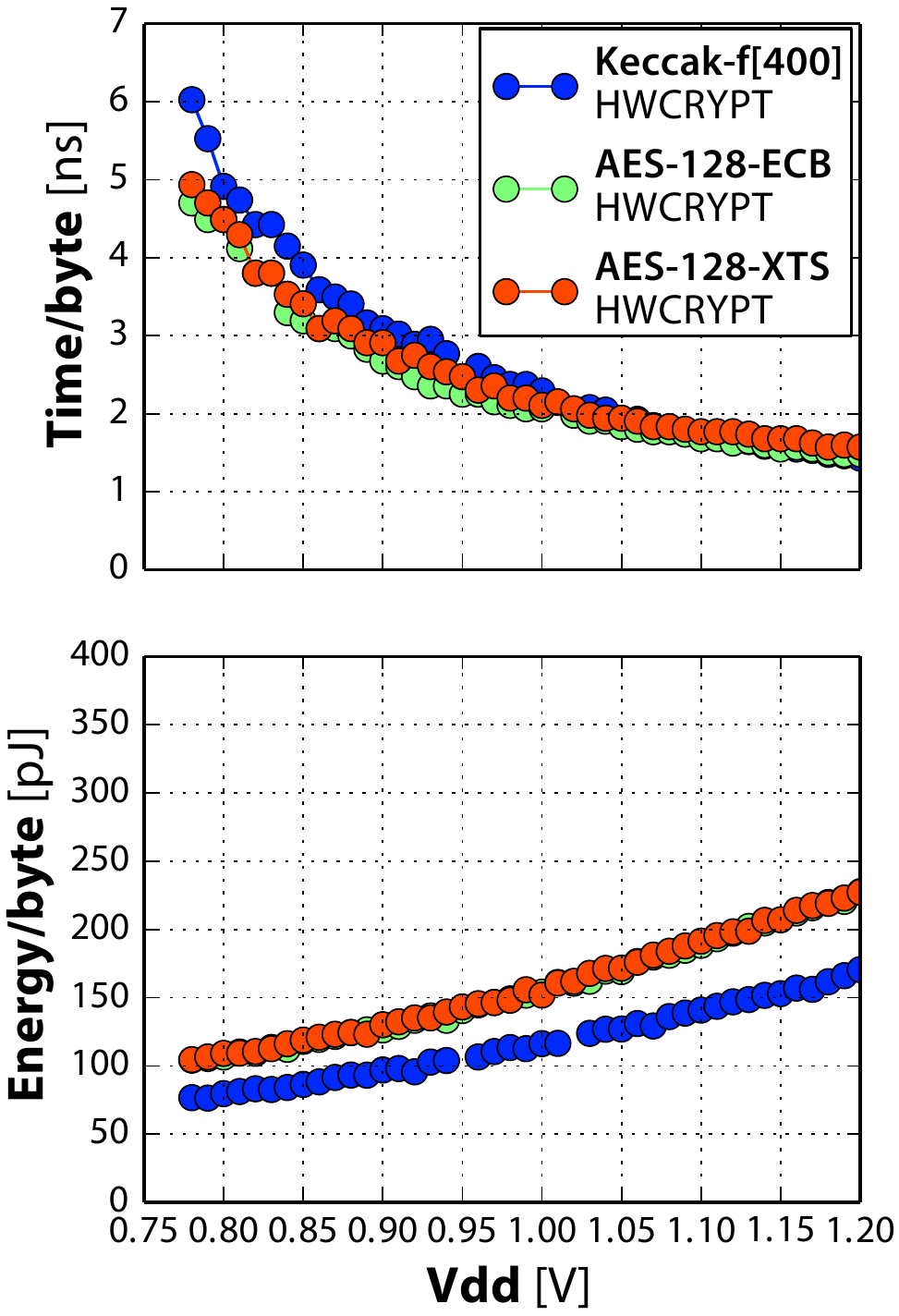}
  }
  \;
  \subfloat[HWCE time/energy per pixel.\label{fig:hwce_cpx_eff}] {
    \centering
    \includegraphics[clip, trim= 0pt 0pt 0pt 0pt, width=0.22\textwidth]{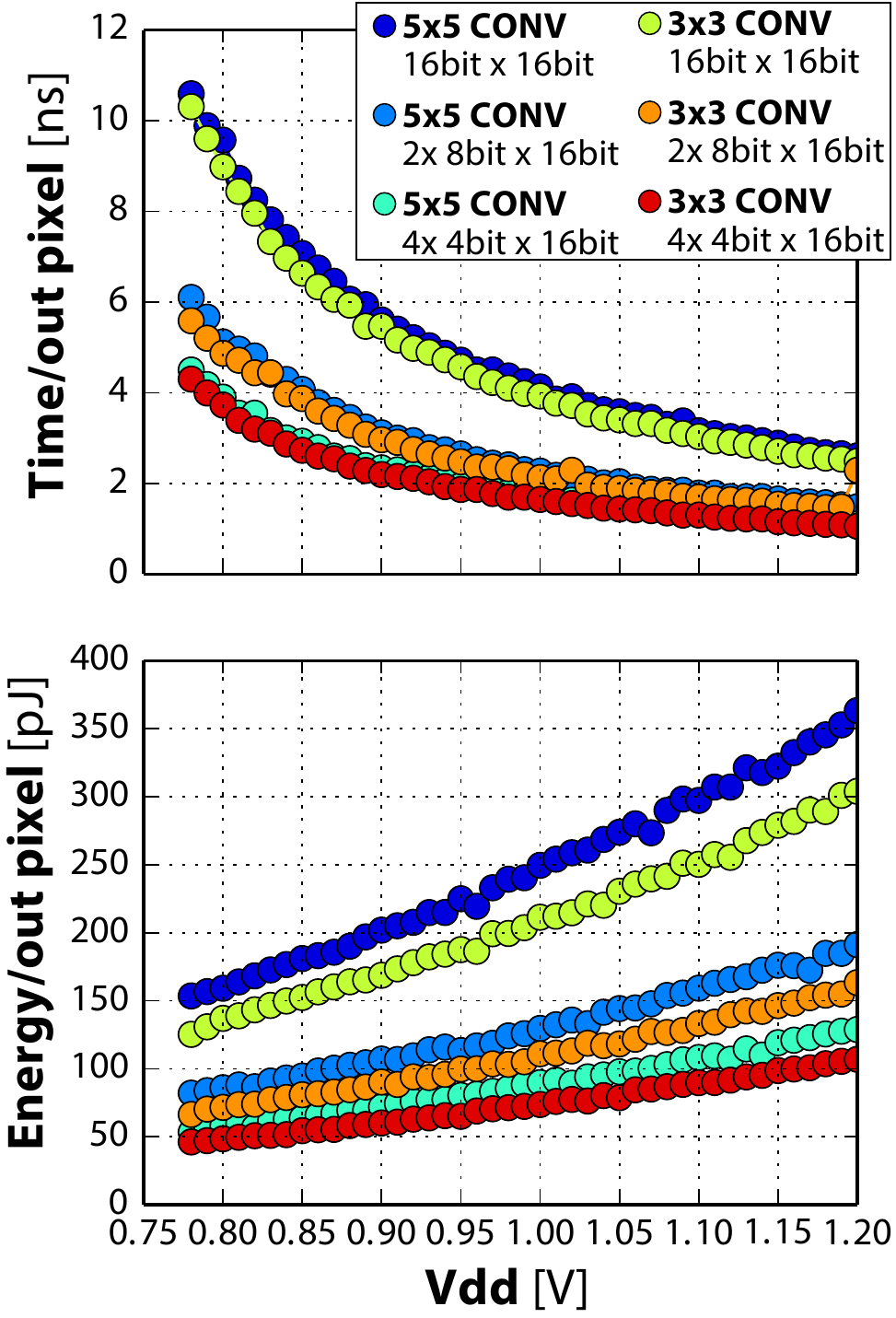}
  }
  \caption{Performance and efficiency of the HWCRYPT and HWCE accelerators in terms of time/energy for elementary output.}
  \label{fig:acc_eff}
\end{figure}

\begingroup
\linespread{0.8}
\renewcommand*{\thefootnote}{\alph{footnote}}
\begin{table*}[tbh]\scriptsize
\centering
  \begin{threeparttable}
    \begin{savenotes}
    \begin{tabulary}{\textwidth}{C l C C C C C C C C C C}
                                                           & & \multirow{2}{*}{\textit{Technology}}              & \multirow{2}{*}{\textit{Area}}  & \multirow{2}{*}{\textit{Power}\footnotemark[1]{}} & \textit{Conv. Perf.\footnotemark[2]{}} & \multirow{2}{*}{\textit{Conv. Eff.\footnotemark[2]{}}} & \textit{Enc. Perf.\footnotemark[3]{}} & \textit{Enc. Eff.\footnotemark[3]{}} & \textit{SW Perf.} & \multirow{2}{*}{\textit{SW Eff.}} & \textit{Eq. Eff.\footnotemark[4]{}} \\  %
                                                                                 &           &                   & [mm${}^2$]                      & [mW]                   & [GMAC/s]       & [GMAC/s/W]          & [Gbit/s]               & [Gbit/s/W]    & [MIPS]    & [MIPS/mW] & [pJ/op] \\
    \toprule
    \multirow{4}{*}{\rot{\textit{AES}}}
    & {Mathew et al.\,\cite{mathew_340_2015} @ 0.43V, 324MHz}                                 & { Intel\,22nm }  & 2.74$\times$10${}^{-3}$    & 0.43                   & -              & -                   & 0.124                  & 289           & -         & -         & 0.19\footnotemark[7]{} \\
    & {Zhang et al.\,\cite{zhang_compact_2016} @ 0.9V, 1.3GHz}                                & { TSMC\,40nm }   & 4.29$\times$10${}^{-3}$         & 4.39                   & -              & -                   & 0.446                  & 113           & -         & -         & 0.49\footnotemark[7]{} \\
    & {Zhao et al.\,\cite{zhao_novel_2015}  @ 0.5V, 34MHz}                                    & { 65nm\,LL }     & 0.013    & 0.05                   & -              & -                   & 0.027                  & 574           & -         & -         & 0.10\footnotemark[7]{} \\
    & {Hocquet et al.\,\cite{hocquet_harvesting_2011} @ 0.36V, 0.32MHz}                       & { 65nm\,LP }     & 0.018    & 2.5$\times$10${}^{-4}$ & -              & -                   & 3.6$\times$10${}^{-7}$ & 144           & -         & -         & 0.39\footnotemark[7]{} \\
    \midrule
    \multirow{5}{*}{\rot{\textit{CNN}}}
    & {Origami\,\cite{cavigelli_803_2015} @ 0.8V, 190MHz}                                     & { UMC\,65nm}     & 3.09        & 93                     & 37             & 402                 & -                      & -             & -         & -         & 0.69\footnotemark[7]{} \\
    & {ShiDianNao\,\cite{ShiDianNao}}                                                         & { 65nm}          & 4.86   & 320                    & 64             & 200                 & -                      & -             & -         & -         & 1.39\footnotemark[7]{} \\
    & {Eyeriss\,\cite{EyerissChen2016} @ 1V, 200MHz}                                          & { TSMC\,65nm\,LP}& 12.25          & 278                    & 23             & 83                  & -                      & -             & -         & -         & 3.35\footnotemark[7]{} \\
    & {Jaehyeong~et~al.\,\cite{JaehyeongISSCC2016} @ 1.2V, 125MHz }                           & { 65nm}          & 16.00   & 45\footnotemark[5]     & 32             & 710\footnotemark[5] & -                      & -             & -         & -         & 0.39\footnotemark[7]{} \\
    & {Park~et~al.\,\cite{park_1.93tops/w_2016} @ 1.2V, 200MHz}                               & { 65nm}          & 10.00   & 37\footnotemark[6]     & 41             & 1108\footnotemark[6]& -                      & -             & -         & -         & 0.25\footnotemark[7]{} \\
    \midrule
    \multirow{4}{*}{\rot{\textit{IoT}}}
    & {SleepWalker\,\cite{bol_sleepwalker_2013} @ 0.4V, 25MHz}                                & { 65nm}          & 0.42   & 0.175                  & -              & -                   & -                      & -             & 25        & 143       & 6.99 \\
    & {Myers~et~al.\,\cite{MyersISSCC2015} @ 0.4V, 0.7MHz}                                    & { 65nm}          & 3.76   & 0.008                  & -              & -                   & -                      & -             & 0.7       & 88        & 11.4 \\
    & {Konijnenburg~et~al.\,\cite{konijnenburg_multi_2016} @ 1.2V, 10MHz   }                  & { 180nm}         & 37.7   & 0.52                   & -              & -                   & -                      & -             & 10.4      & 20        & 50.0 \\
    & {Mia~Wallace\,\cite{pullini_heterogeneous_2016} @ 0.65V, 68MHz }                        & { UMC\,65nm }    & 7.4        & 9.2                    & 2.41           & 261                 & -                      & -             & 270       & 29        & 22.5 \\
    \midrule
    \multirow{3}{*}{\rot{\textit{\textbf{Fulmine}}}}
    & \textbf{\textsc{cry-cnn-sw} @ 0.8V, 85MHz}  & \multirow{3}{*}{\textbf{UMC\,65nm\,LL}}   & \multirow{3}{*}{\textbf{6.86}}  & \textbf{24       }     & \textbf{4.64}  & \textbf{309}          & \textbf{1.78}          & \textbf{67 }  & \textbf{333}    &    \textbf{14}          & \multirow{3}{*}{\textbf{5.74}} \\
    & \textbf{\textsc{kec-cnn-sw} @ 0.8V, 104MHz} &                                    &   & \textbf{13       }     & \textbf{6.35}  & \textbf{465}          & \textbf{1.6 }          & \textbf{100}  & \textbf{408}    &    \textbf{31}          & \\
    & \textbf{\textsc{sw} @ 0.8V, 120MHz}         &                                    &   & \textbf{12       }     & -              & -                     & -                      & -             & \textbf{470}    &    \textbf{39}          & \\
    \bottomrule
    \end{tabulary}
    \end{savenotes}
    \begin{tablenotes}
      \linespread{0.9}
      \item \footnotemark[1]{} Power and efficiency numbers refer to core power, excluding I/Os.
      \item \footnotemark[2]{} Considering 1 MAC = 2 ops where Gop/s are reported. \textit{Fulmine} numbers refer to the 4bit weights mode.
      \item \footnotemark[3]{} Refers to AES-128-$\{$ECB,XTS$\}$ for \textit{Fulmine} in \textsc{cry-cnn-sw}; \keccak{} for \textit{Fulmine} in \textsc{kec-cnn-sw}; else AES-128-ECB.
      \item \footnotemark[4]{} Considering the local face detection workload of Section \ref{sec:results_face}. 1op = 1 OpenRISC equivalent instruction from the set defined in \cite{_openrisc_2012}.
      \item \footnotemark[5]{} Weights produced on-chip from a small set of PCA bases to save area/power. No evaluation on the general validity of this approach is presented in \cite{JaehyeongISSCC2016}.
      \item \footnotemark[6]{} Performance \& power of inference engines only, estimating they are responsible for 20\% of total power.
      \item \footnotemark[7]{} ASIC equivalent efficiency refers to an AES-only or CNN-only equivalent workload.
    \end{tablenotes}
  \end{threeparttable}
\caption{
  Comparison between \textit{Fulmine} and several platforms representative of the state-of-the-art in encryption, data analytics, and IoT end-nodes.
}
\label{tab:comparison}
\end{table*}
\linespread{1}
\endgroup

In this Section, we analyze measured performance and efficiency of our platform on the manufactured \textit{Fulmine} prototype chips, fabricated in UMC 65\,nm LL 1P8M technology\correction{, in a \SI{2.62}{\milli\meter}$\times$\SI{2.62}{\milli\meter} die}.
\crossout{
Figure~\ref{fig:chip_micrograph} shows a microphotograph of a manufactured \textit{Fulmine} chip, which occupies an area of \SI{2.62}{\milli\meter}$\times$\SI{2.62}{\milli\meter}.
}

\subsection{System-on-Chip Operating Modes}
\label{sec:op_modes}

An important constraint for the design of small, deeply embedded systems such as the \textit{Fulmine} SoC is the maximum supported power envelope.
This parameter is important to select the system battery and the external DC/DC converter.
To maximize energy efficiency, the worst case for the DC/DC converter (i.e. the peak power) should not be too far from the average working power to be delivered.
However, a SoC like \textit{Fulmine} can operate in many different conditions: in pure software, with part of the accelerator functionality available, or with both accelerators available.
These modes are characterized by very different average switching activities and active power consumption.
\crossout{
During the design of the \textit{Fulmine} SoC, three distinct operating modes were defined, enabling to choose the right combination of active cores and accelerators in an application-driven fashion.
When an application is run entirely in software, it is often useful to push frequency as much as possible to maintain a performance constraint.
Conversely, execution on accelerator cores is orders of magnitude faster than software, and it can be executed at a relaxed operating frequency with less overall cluster power consumption.
}
\correction{
In pure software mode, it is often desirable to push frequency as much as possible, while when using accelerators it can be convenient to relax it to improve power consumption.
}
Moreover, some of the internal accelerator datapaths are not easily pipelined, as adding pipeline stages severely hits throughput - this is the case of the HWCRYPT sponge engine (Section~\ref{sec:hwcrypt}), which relies on tight loops of \keccak{} rounds as visible in the datapath in Figure~\ref{fig:fulmine_hwcrypt_dp}.
Relaxing these paths can improve the overall synthesis results for the rest of the circuit.

Multi-corner multi-mode synthesis and place \& route were used to define three operating modes \correction{that the developer can statically select for the target application}:
in the \textsc{cry-cnn-sw} mode, all accelerators and cores can be used.
In the \textsc{kec-cnn-sw} mode, cores and part of the accelerators can be used: the HWCE fully, the HWCRYPT limited to \keccak{} primitives.
In this mode, the frequency can be pushed significantly further than in the \textsc{cry-cnn-sw} mode.
Finally, in the \textsc{sw} mode, only the cores are active, and the operating frequency can be maximized.
Figure~\ref{fig:cluster_op} shows frequency scaling in the three operating modes while varying the cluster operating voltage $V_{DD}$.
The three modes were designed so that at $V_{DD}=$\SI{1.2}{\volt}, current consumption under full load is close to \SI{100}{\milli\ampere} (i.e. \SI{120}{\milli\watt} of power consumption), as can be seen in Figure~\ref{fig:cluster_pow}.

\subsection{HWCRYPT Performance and Power Evaluation}
\label{sec:results_hwcrypt}

Due to a throughput oriented hardware implementation, HWCRYPT achieves a significant acceleration compared to an optimized software implementation running on the OpenRISC cores.
To encrypt one 8\,kB block of data using the AES-128-ECB mode, HWCRYPT requires $\sim$3100 clock cycles including the initial configuration of the accelerator.
This is a 450$\times$ speedup compared to a software implementation on one core.
When parallelizing the software implementation to all four cores, the hardware accelerator still reaches a speedup of 120$\times$.
The throughput of HWCRYPT in AES-128-ECB mode is 0.38 cycles per byte~(cpb).

The performance of the AES-128-XTS mode is the same with respect to the ECB mode, thanks to parallel tweak computation and encryption.
When comparing that to an optimized software implementation on a single core, this speeds up the throughput by a factor of 495$\times$ and by a factor 287$\times$ when running on four cores.
It is important to note that, contrarily to the ECB mode, XTS encryption cannot be efficiently parallelized in software due to a data dependency during the tweak computation step.

The authenticated encryption scheme based on \keccak{} achieves a throughput of 0.51\,cpb by utilizing both permutation instances in parallel.
The first permutation encrypts the data and the second one is used to compute the message authentication code to provide integrity and authenticity.
This performance is achieved in a maximum-rate configuration of 128\,bit per permutation call and 20 rounds as specified by \keccak{}.
Reducing the rate and/or increasing the number of invoked permuations decreases the throughput while increasing the security margin.

In Figure~\ref{fig:hwcrypt_cpb_eff}, we present the performance of HWCRYPT in terms of time and energy per byte, while scaling the $V_{DD}$ operating voltage of the cluster. When normalizing these values to the power consumption, we reach a performance of 67\,Gbit/s/W for AES-128-XTS and 100\,Gbit/s/W for \keccak{}-based authenticated encryption respectively.

\subsection{HWCE Performance and Power Evaluation}
\label{sec:results_hwce}

The \textit{Fulmine} SoC includes many distinct ways to perform the basic operation of CNNs, i.e. 2D convolutions.
In software, a na\"{i}ve single core implementation of a 5$\times$5 convolution filter has a throughput of 94\,cycles per pixel.
Parallel execution on four cores can provide almost ideal speedup reaching 24\,cycles/px.
Thanks to the SIMD extensions described in Section~\ref{sec:archi}, an optimized multi-core version can be sped up by almost 2$\times$ down to 13\,cycles/px on average.

With respect to this baseline, the HWCE can provide a significant additional speedup by employing its parallel datapath, the line buffer (which saves input data fetch memory bandwidth), and weight precision scaling.
We measured average throughput by running a full-platform benchmark, which therefore takes into account the overheads for real world usage: line buffer fill time, memory contention from cores, self-contention by HWCE inputs/outputs trying to access the same TCDM bank in a given cycle.
Considering the full precision 16\,bit mode for the weights, we measured an average inverse throughput of 1.14\,cycles per output pixel for 5$\times$5 convolutions and 1.07\,cycles per output pixel for 3$\times$3 convolutions - the two sizes directly supported by the internal datapath of the HWCE.
This is equivalent to a 82$\times$ speedup with respect to the na\"{i}ve single core baseline, or 11$\times$ with respect to a fully optimized 4-core version.

As described in Section~\ref{sec:hwce}, the HWCE datapath enables application-driven scaling of arithmetic precision in exchange for higher throughput and energy efficiency.
In the 8\,bit precision mode, average inverse throughput is scaled to 0.61\,cycles/px and 0.58\,cycles/px for the 5$\times$5 and 3$\times$3 filters, respectively; in 4bit mode, this is further improved to 0.45\,cycles/px and 0.43\, cycles/px, respectively.
In the 4\,bit precision mode, the HWCE is fully using its 4-port memory bandwidth towards the TCDM in order to load 4 $\mathbf{y}_\mathrm{in}$ partial results and store back 4 $\mathbf{y}_\mathrm{out}$ ones.
Further performance scaling would therefore require an increase in memory bandwidth.%
Figure~\ref{fig:hwce_cpx_eff} reports time and energy per pixel, running the same set of filters in the \textsc{kec-cnn-sw} operating mode while scaling the $V_{DD}$ operating voltage.
At \SI{0.8}{\volt}, the energy to spend for an output pixel can be as low as \SI{50}{\pico\joule} per pixel, equivalent to 465\,GMAC/s/W for a 5$\times$5 filter.

\subsection{Comparison with State-of-the-Art}

\crossout{To compare special-purpose devices with general-purpose microcontrollers, }
Table \ref{tab:comparison} compares \textit{Fulmine} with the architectures that define the boundaries of the secure data analytics application space \correction{described in Section \ref{sec:related}}.
Apart from area, power and performance, we also use an \textit{equivalent energy efficiency} metric defined as the energy that a platform has to spend to perform an elementary RISC operation\footnote{This is computed as \crossout{the total energy to execute a given workload in a platform by the number of equivalent OpenRISC 1200 instructions; specifically, we used}\correction{the total energy per instruction on} the workload presented in Section \ref{sec:results_face}, which provides a balanced mix of encryption, convolution, other SW-based filters.}.
\textit{Fulmine} achieves the highest result on this metric, \SI{5.74}{\pico\joule} per operation, thanks to the cooperation between its three kinds of processing engines.
The second-best result is of SleepWalker (\SI{6.99}{\pico\joule}) - but in an operating point where execution takes 89$\times$ more time than in the case of \textit{Fulmine}.

Moreover, \textit{Fulmine} provides better area efficiency than what is available in other IoT end-nodes:
\crossout{
For example, even discounting all complexity related to a setup with many computing chips on the same board, 32 SleepWalker chips would be needed to achieve a performance similar to ours on the  face detection workload of Section \ref{sec:results_face}.
}
\correction{32 SleepWalker chips would be needed to achieve the same performance as Fulmine in the workload of Section \ref{sec:results_face}.}
On the other hand, \crossout{while} coupling an efficient IoT microcontroller with external accelerators can theoretically provide an effective solution, \crossout{this would require continued data exchange between the devices for most of the secure data analytics scenarios for IoT devices}\correction{but it requires continuous high-bandwidth data exchange from chip-to-chip, which is typically not practical in low-power systems}.
\crossout{The necessity for high bandwidth for chip-to-chip communication within a ultra-low-power system makes this solution typically impractical.}
\correction{
Conversely, in Fulmine HW accelerators are coupled to the cluster cores via the shared L1 memory, and no copy at all is required - only a simple pointer exchange.
}

\correction{
For IoT endnodes, the smaller footprint of a System-on-Chip solution can also provide an advantage with respect to a traditional system on board, which is heavier and bulkier.
Taking this reasoning one step further, while it is not always possible to place sensors and computing logic on the same die, the system we propose could be coupled to a sensor in a System-on-Package solution, requiring only a single die-to-die connection.
Competing systems listed in Table \ref{tab:comparison} would require the integration of more than two dies on the same package, resulting in a more complex and expensive design.
}
\crossout{
Scaling to more deeply integrated technology and/or operating voltage could provide very significant benefits in terms of energy.
If we consider a 28\,nm technology at $V_{DD}=$\SI{0.6}{\volt} with a similar frequency target, power would scale to $\sim$\SI{4}{\milli\watt}, with an overall improvement to energy efficiency by 6$\times$, making it comparable to dedicated accelerators and enabling even more complex integrated workloads.
}

\section{Use cases}
\label{sec:results_usecases}

\begin{figure}[b]
    \centering
    \includegraphics[clip, trim= 0pt 0pt 0pt 0pt, width=0.32\textwidth]{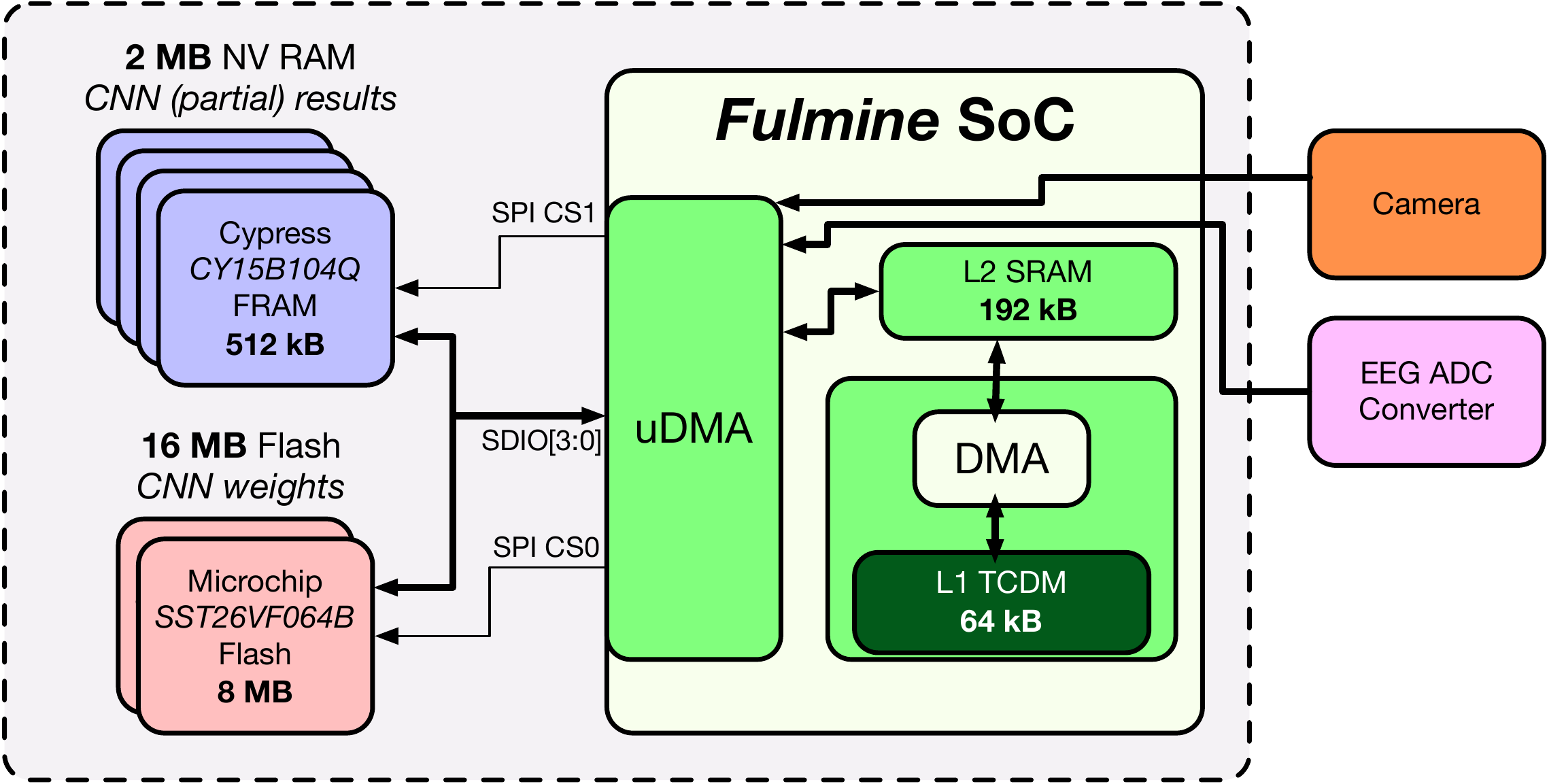}
    \caption{A \textit{Fulmine} SoC connected to 16\,MB of Flash, 2\,MB of FRAM, and sensors (the grey area is taken into account for power estimations).}
    \label{fig:setup}
\end{figure}

\crossout{
To evaluate the \textit{Fulmine} SoC in full end-to-end applications, we consider three distinct use cases, representative of a wider set of possible applications for secure analytics in the IoT domain.
We consider secure surveillance on a fully autonomous aerial vehicle based on a complex state-of-the-art CNN (\textit{ResNet-20}~\cite{he_deep_2015}); a smartwatch-based secure authentication using local face detection and remote cloud-based recognition, based on a smaller CNN~\cite{li_convolutional_2015}; and seizure detection based on principal component analysis and wavelet transform~\cite{benatti_scalable_2016}, with secure data collection for further analysis.
The proposed use cases are not intended to exhaust the space for applications of the presented SoC, but to explain how the architectural features of \textit{Fulmine} can be used to make advanced secure analytics capabilities feasible in small low-power computing end-nodes.
}
\correction{
To evaluate the \textit{Fulmine} SoC in full end-to-end applications, we propose three distinct use cases, which represent a necessarily incomplete selection of possible security- and performance-critical IoT sensor analytics applications.
The first use case represents deep-learning based sensor analytics workloads that are predominantly executed locally on the endnode, but require security to access unsafe external memory (secure autonomous aerial surveillance, Section \ref{sec:results_uav}); the second one represents workloads executed only in part on the endnode, which therefore require secured connectivity with an external server (local face detection and remote recognition, Section \ref{sec:results_face}).
Finally, the third use case represents workloads in which, while analytics is performed online, data must also be collected for longer term monitoring (seizure detection and monitoring, Section \ref{sec:results_eeg}).
}

For our evaluation, we consider the system shown in Figure~\ref{fig:setup}.
\crossout{
To be able to work on state-of-the-art analytics based on deep learning, the \textit{Fulmine} SoC is connected to low-power memory and Flash storage.
}
We use two banks (16\,MB) of Microchip SST26VF064\,bit quad-SPI flash memory to host the weights for a deep CNN as \textit{ResNet-20}; each bank consumes down to \SI{15}{\micro\ampere} in standby and a maximum of \SI{15}{\milli\ampere}@\SI{3.6}{\volt} in QPI mode.
Moreover, we use 2\,MB of non-volatile Cypress CY15B104Q ferroelectric RAM (FRAM) as a temporary memory for partial results.
Four banks are connected in a bit-interleaved fashion to allow access with quad-SPI bandwidth.
Both the FRAM and the flash, as well as a camera and an ADC input, are connected to the \textit{Fulmine} uDMA, which can be used to transfer data to/from the SoC L2 memory.
The cluster then transfers tiles of the input data to operate on and writes results back to L2 via DMA transfers.
\correction{
We focus on the power spent for actual computation rather than on system power, i.e. we include power spent in transfers from memory used during the computation, but exclude data acquisition/transmission\footnotemark.
}

\crossout{
We measured performance on the \textit{Fulmine} SoC for the kernels composing the three applications and for SPI and DMA transfers via RTL simulation, and the relative power consumption of the \textsc{soc} and \textsc{cluster} power domains by direct measurement on the fabricated chip, using an Advantest SoCV93000 integrated circuit tester.
We focus on the power spent for actual computation rather than on system power.
We intend this in a broad sense, including e.g. power spent in memory transfers to/from flash and FRAM that are necessary for the computation itself, but excluding power spent for data acquisition and transmission, which are clearly separated from the computation phase.
For the two external memories, we used publicly available data from their datasheets, considering worst case power whenever appropriate.
}
\footnotetext{
\correction{
We measured performance on each kernel composing the three applications and for SPI and DMA transfers via RTL simulation, and the related power consumption by direct measurement using an Advantest SoCV93000 integrated circuit tester, encapsulating the target kernel within an infinite loop.
Power is measured at two distinct frequencies to obtain leakage and dynamic power density via linear regression.
For external memories, we used publicly available data from their datasheets, considering always the worst case.
}
}

\subsection{Secure Autonomous Aerial Surveillance}
\label{sec:results_uav}

\begin{figure*}[tb]
    \centering
    \includegraphics[clip, trim= 0pt 0pt 0pt 0pt, width=0.9\textwidth]{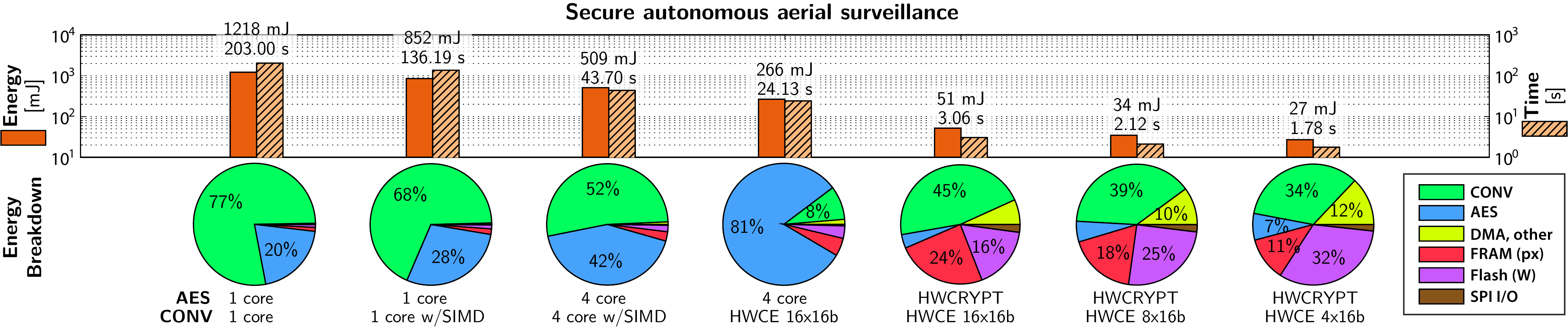}
    \caption{Secure autonomous aerial surveillance use case based on a \textit{ResNet-20} CNN~\cite{he_deep_2015} with AES-128-XTS encryption for all weights and partial results. \textsc{kec-cnn-sw} and \textsc{cry-cnn-sw} operating modes at $V_{DD}=$\SI{0.8}{\volt}.}
    \label{fig:resnet_20}
\end{figure*}

For the secure autonomous aerial surveillance use case, we consider deploying the system of Figure~\ref{fig:setup} on a low-power nano-UAV such as a CrazyFlie nano quadcopter~\cite{crazyflie_nano_2016}.
Storms of tens or hundreds of these devices could provide diffused, fully autonomous, and low energy footprint aerial surveillance.
\crossout{
Continuous data transmission from onboard cameras to a centralized surveillance server for analysis would require significant power from the limited UAV battery and reduce the overall flight time.
Moreover, being fully dependent on wireless transmission might be detrimental from the UAV reliability viewpoint, especially in situations such as disaster recovery where UAV storms might otherwise be used with success.
A sensible choice is therefore to perform part of the surveillance analysis directly on the UAVs, and then send it to the collecting server in the cloud.
This saves energy and increases reliability, as collected labels are synthetic and can be collected for some time on the device if the wireless connection is temporarily unavailable.
The power budget for computing on this category of aerial vehicle is extremely limited, as more than 90\% of the battery must be dedicated to the quadrotor engines.
}
\correction{
In these vehicles the power budget for computing is extremely limited (more than 90\% of the battery must be dedicated to the quadrotor engines), and continuous wireless data transmission from cameras is not an option due to its power overhead.
Local elaboration and transmission of high-level labeling information provides a more efficient usage of the available power, while also granting greater availability in situations like disaster relief, where wireless propagation might be non-ideal and enable only low-bandwidth communication.
}

\crossout{
An additional constraint if one wants to use deep CNNs of medium and large size is that they require external memory for storage of weights and partial results.
}
\correction{
Deployment of state-of-the-art deep CNNs on these devices naturally requires external memory for the storage of weights and partial results.
}
These memories cannot be considered to be secure\correction{, as}
\crossout{First, }the weights deployed in the Flash\crossout{ memory can be considered}\correction{are} an important intellectual property\crossout{ and therefore need to be protected. %
Second, as the} and UAVs are fully autonomous, \crossout{they are potentially}\correction{therefore} vulnerable to malicious physical hijacking . %
Partial results stored in the FRAM and SPI traffic could be monitored or modified by an external agent, with the purpose of changing the final result classified by the UAV.
Strong encryption for weights and partial results can significantly alleviate this issue, at the cost of a huge overhead on top of the pure data analytics workload.

\crossout{To model this use case,}\correction{Here} we consider a deep \textit{ResNet-20} CNN~\cite{he_deep_2015} to classify scenes captured from a low power sensor producing a 224$\times$224 input image.
\textit{ResNet-20} has been shown to be effective on CIFAR-10 classification but can also be trained for other complex tasks\correction{, and it is in general a good representative of state-of-the-art CNNs of medium size}.
It consists of more than $1.35\times 10^9$ operations, a considerable workload for a low power end-node.
\crossout{It also requires using}External memories \correction{are required} for both weights (with a footprint of 8.9\,MB considering 16 bits of precision) and partial results (with a maximum footprint of 1.5\,MB for the output of the first layer).
\crossout{On top of this, }All weights and partial results are en-/decrypted with AES-128-XTS; the \textit{Fulmine} cluster is considered the only secure enclave in which decrypted data can reside.

Figure~\ref{fig:resnet_20} shows execution time and energy spent at \SI{0.8}{\volt} for this compound workload.
We exploit the fast frequency switching capabilities of \textit{Fulmine} \crossout{described in Section~\ref{sec:archi}} to dynamically switch from the \textsc{cry-cnn-sw} operating mode (at \SI{85}{\mega\hertz}) when executing AES to the \textsc{kec-cnn-sw} operating mode (at \SI{104}{\mega\hertz}) when executing other kernels.
The figure also shows a breakdown of energy consumption regarding kernels (convolution \correction{\textsc{CONV}}, encryption \correction{\textsc{AES}}), \crossout{other components of the CNN} \correction{densely connected CNN layers (\textsc{DENSE})}, DMA transfers \correction{and other parts of the CNN (\textsc{DMA, other})}, and external memories\correction{ and I/O (\textsc{FRAM}, \textsc{Flash}, \textsc{SPI I/O})}.
In the baseline, where all the workload is run in software on a single core, energy consumption is entirely dominated by convolutions and encryption, with a 4-to-1 ratio between the two.
When more features of the \textit{Fulmine} SoC are progressively activated, execution time is decreased by 114$\times$ and energy consumption by 45$\times$, down to \SI{27}{\milli\joule} in total - \SI{3.16}{\pico\joule} per equivalent operation \correction{(defined as an equivalent OpenRISC instruction from \cite{_openrisc_2012})}.
When CNNs use the HWCE with 4\,bit weights and AES-128-XTS uses the HWCRYPT, the overall energy breakdown shows that cluster computation is no longer largely dominant, counting for only slightly more than 50\% of the total energy.
\crossout{
The energy expense for weights is decreased by using lower arithmetic accuracy weights, but the FRAM for partial results accounts for more than 30\% of the total energy spent.
}
Additional acceleration would likely require expensive hardware (e.g. more sum-of-products units or more ports in the HWCE) and would \crossout{therefore} yield diminishing returns in terms of energy efficiency.

To concretely estimate whether the results make it feasible to deploy a \textit{ResNet-20} on a nano-UAV, consider that a CrazyFlie UAV~\cite{crazyflie_nano_2016} can fly for up to 7 minutes.
Continuous execution of secure \textit{ResNet-20} during this flight time corresponds to a total of 235 iterations in the operating point considered here.
This would consume a total of \SI{6.4}{\joule} of energy - less than 0.25\% of the \SI{2590}{\joule} available in the onboard battery\correction{ - and the low peak power of \SI{24}{\milli\watt} makes this concretely achievable in an autonomous device}.
\crossout{
Together with peak power consumption, which is less than \SI{24}{\milli\watt}, this result constitutes a strong argument in favor of the effectiveness of the platform we propose for complex secure classification workloads executed directly onboard of a surveillance device of this kind.
}

\subsection{Local Face Detection with Secured Remote Recognition}
\label{sec:results_face}

\begin{figure*}[tb]
    \centering
    \includegraphics[clip, trim= 0pt 0pt 0pt 0pt, width=0.9\textwidth]{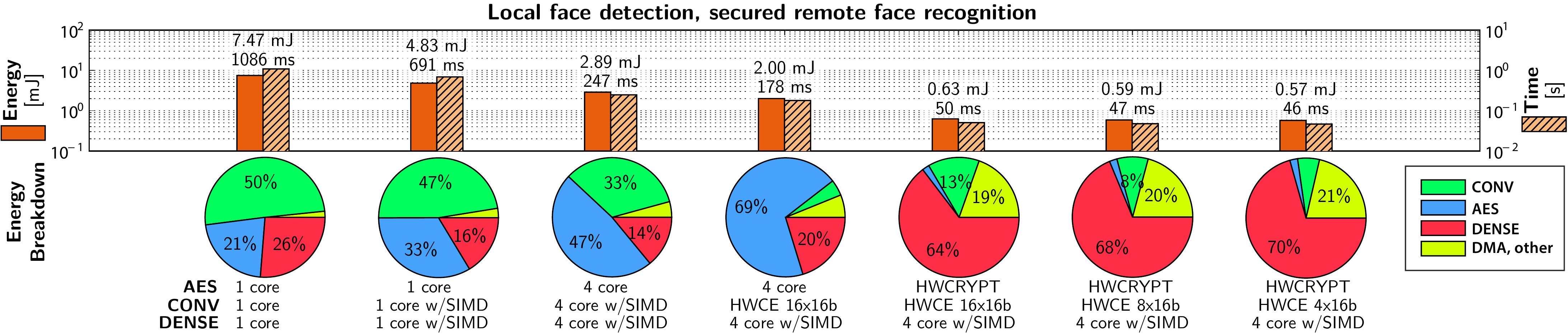}
    \caption{Local face detection, secured remote recognition use case based on the 12-net and 24-net CNNs from Li et al.~\cite{li_convolutional_2015} on a 224$\times$224 input image, with full AES-128-XTS encryption of the image if a potential face is detected. \textsc{cry-cnn-sw} operating mode at $V_{DD}=$\SI{0.8}{\volt}. We consider that the first stage 12-net classifies 10\% of the input image as containing faces, and that the second stage 24-net is applied only to that fraction.}
    \label{fig:cvpr15li_24net}
\end{figure*}

Complete on-device computation might not be the most advantageous approach for all applications, particularly for those that can be clearly divided in a lower effort \emph{triggering} stage and a higher effort one that is only seldom executed.
A good example is the problem of face recognition.
While state-of-the-art face recognition requires a significant workload in the order of billions of operations (e.g. FaceNet~\cite{schroff_facenet:_2015}), the problem can be easily decomposed in two stages: one where the input image is scanned to detect the presence of a face, and another where the detected faces are recognized.
The first stage could be run continuously on a low-power wearable device such as a smartwatch, using an external device (e.g. a smartphone, the cloud) to compute the much rarer and much more complex second stage.

\crossout{In this use case, }We envision \textit{Fulmine} to be integrated into an ultra-low power~(ULP) smartwatch platform similar to that presented in Conti et al.~\cite{conti_accelerated_2016}.
We consider a similar camera with the one used in Section~\ref{sec:results_uav} producing a 224$\times$224 input image.
Face detection is performed locally, using the first two stages (12-net and 24-net) of the multi-stage CNN proposed by Li et al.~\cite{li_convolutional_2015}.
If faces are detected by this two-stage CNN, the full input image is encrypted and transferred to a coupled smartphone for the recognition phase.
The networks are applied to small separate 24$\times$24 windows extracted from the input image; partial results need not be saved from one window to the next. Therefore the CNN does not use any external memory and can rely exclusively on the internal L2.

Figure~\ref{fig:cvpr15li_24net} reports the experimental results for the local face detection use case in terms of energy and execution time.
Baseline energy is almost evenly spent between convolutions, AES-128-XTS encryption, and densely connected CNN layers.
Software optimizations such as parallelization, SIMD extensions are much more effective on convolutional and dense layers than they are on AES, due to \crossout{their highly parallel and regular structure and to} XTS internal data dependencies in the tweak computation.
Using hardware accelerators essentially reduces the energy cost of convolution and on AES-128-XTS to less than 10\% of the total, and leads to a 24$\times$ speedup and a 13$\times$ reduction in energy with respect to the baseline.
\crossout{The final test}\correction{With all optimizations, face detection} takes \SI{0.57}{\milli\joule} or \SI{5.74}{\pico\joule} per elementary operation.
\crossout{
Further reduction could be enabled by algorithmic changes that favor a deeper network with more convolutional layers to one with many densely connected layers.
}
\crossout{
Even without these potential improvements, the result we show enables deployment of this algorithm on a real ULP smartwatch.
If we consider it to be powered by a small lithium-ion polymer \SI{4}{\volt} \SI{150}{\milli\ampere\hour} battery,
}
\correction{This} face detection could be performed with no interruption for roughly 1.6 days before exhausting the battery charge\correction{, if we consider a small \SI{4}{\volt} \SI{150}{\milli\ampere\hour} lithium-ion polymer battery}.
Duty cycling, taking advantage of the power management features of the SoC described in Section~\ref{sec:archi}, can prolong this time considerably.

\subsection{Seizure Detection and Secure Long-Term Monitoring}
\label{sec:results_eeg}

\begin{figure}[tb]
    \centering
    \includegraphics[clip, trim= 0pt 0pt 0pt 0pt, width=0.48\textwidth]{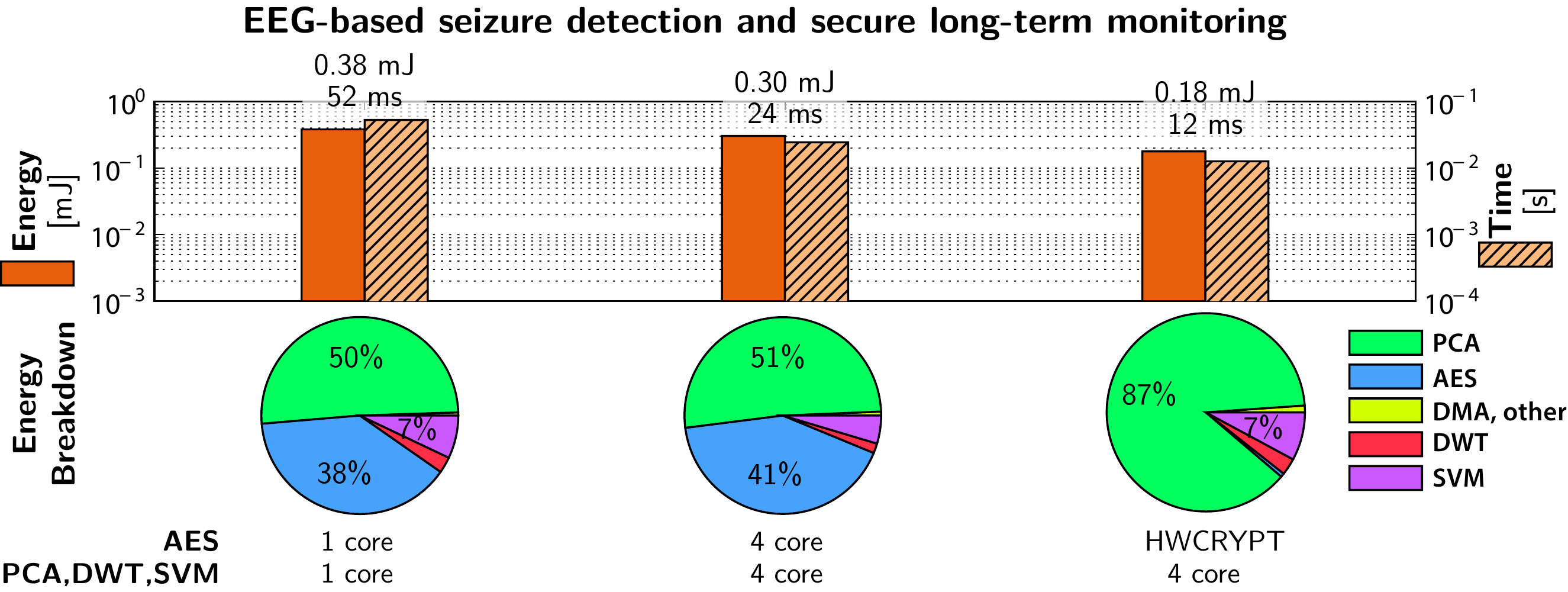}
    \caption{EEG-based seizure detection and secure data collection. \textsc{cry-cnn-sw} operating mode at $V_{DD}=$\SI{0.8}{\volt}.}
    \label{fig:eeg}
\end{figure}

\crossout{
While the first two use cases we propose focus on vision, that is not the only computationally intensive task that could be executed on a future computing end-node.
}
Extraction of semantically relevant information out of biosignals such as electromyogram (EMG), electrocardiogram (ECG), and electroencephalogram (EEG) is a potentially huge market for low-power footprint IoT devices\crossout{, as an enabler technology for smart fitness, personalized healthcare, and other human augmentation technologies}.
\crossout{In this use case,}\correction{Here} we consider a seizure detection \crossout{healthcare} application based on a support vector machine~(SVM) trained on energy coefficients extracted from the principal components \correction{analysis~(PCA)} of a multi-channel EEG signal~\cite{benatti_versatile_2015}\cite{benatti_scalable_2016}.
The sampling frequency is \SI{256}{\hertz} with 50\% overlapped windows, i.e. seizure detection is performed every \SI{0.5}{\second}.
Starting from a 256-sample window of 23 input EEG channels (represented as 32\,bit fixed-point numbers), PCA is applied to extract 9 components, that are then transformed \crossout{in a wavelet representation} by a digital wavelet transform (DWT) \correction{to extract energy coefficients, which are classified by an SVM}.
\crossout{Energy coefficients are extracted from this representation.
Finally, a classifier SVM is used to determine if there is a seizure.}
For long-term monitoring, the components produced by the PCA have to be collected and sent to the network to be stored or analyzed\correction{, which requires encryption due to the sensitivity of this data}.
\crossout{
Given the high degree of sensitivity of this data, it cannot be transferred in plain format but must be appropriately encrypted, for which we consider AES-128-XTS.
}

Figure~\ref{fig:eeg} shows the results in terms of energy (split down between the various kernels) and execution time.
Several components of PCA, like diagonalization, are not amenable to parallelization.
Nonetheless, we observe a 2.6$\times$ speedup with four cores excluding AES encryption.
Using the HWCRYPT, encryption becomes a \emph{transparent} step of the algorithm and essentially disappears from the overall energy breakdown.
Therefore, with combined SW parallelization and accelerated encryption, an overall 4.3$\times$ speedup and 2.1$\times$ energy reduction can be achieved.
More importantly, the absolute energy consumption of \SI{0.18}{\milli\joule} (\SI{12.7}{\pico\joule} per operation) means that a typical \SI{2}{\ampere\hour}@\SI{3.3}{\volt} pacemaker battery~\cite{mallela_trends_2004} would suffice for more than 130 million iterations, and more than 750 days if used continuously - as for most of the time the \textit{Fulmine} SoC can be in \crossout{the sleep mode described in Section~\ref{sec:archi}}\correction{\textit{deep sleep} mode}.

\section{Conclusion}
\label{sec:conclusion}

This work presented \textit{Fulmine}, a \SI{65}{\nano\meter} System-on-Chip targeting the emerging class of smart secure near-sensor data analytics for IoT end-nodes.
We achieve this without using aggressive technology or voltage scaling, but through the architectural solution of combining cores and accelerators within a single tightly-coupled cluster.
The use cases we have proposed show that this approach leads to improvements of more than one order of magnitude in time and energy with respect to a pure software based solution, with no sacrifice in terms of flexibility.
The \textit{Fulmine} SoC enables secure, integrated and low-power \textit{secure data analytics} directly within the IoT node.
Without any compromise in terms of security, the proposed SoC enables sensemaking in a budget of a few pJ/op - down to 3.16\,pJ/op in one case, or 315\,Gop/s/W.

\ifCLASSOPTIONcaptionsoff
  \newpage
\fi
\renewcommand{\baselinestretch}{0.95}
\bibliography{./bib/references,./bib/Fulmine_TCAS1_fconti,./bib/Fulmine_TCAS1_minor}

\vspace{-2cm}
\begin{IEEEbiography}[{\includegraphics[width=0.85in,clip,keepaspectratio]{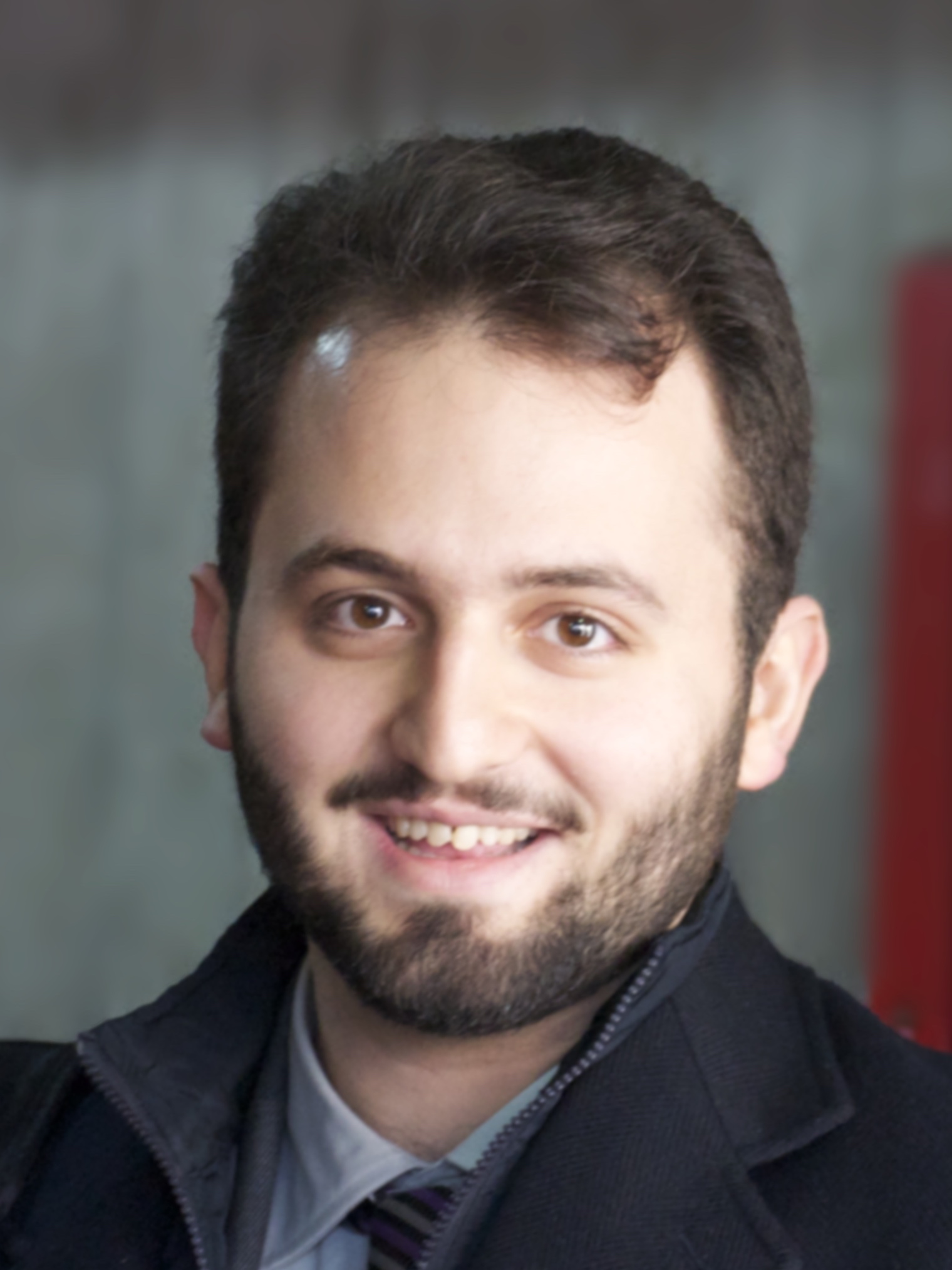}}]{Francesco Conti} received the Ph.D. degree from University of Bologna in 2016 and is currently a post-doctoral researcher at the Integrated Systems Laboratory, ETH Z\"{u}rich, Switzerland and the Energy-Efficient Embedded Systems laboratory, University of Bologna, Italy.
He has co-authored more than 20 papers on international conferences and journals.
His research focuses on energy-efficient multicore architectures and applications of deep learning to low power digital systems.
\end{IEEEbiography}
% \newpage
\vspace{-2cm}
\begin{IEEEbiography}[{\includegraphics[width=0.85in,clip,keepaspectratio]{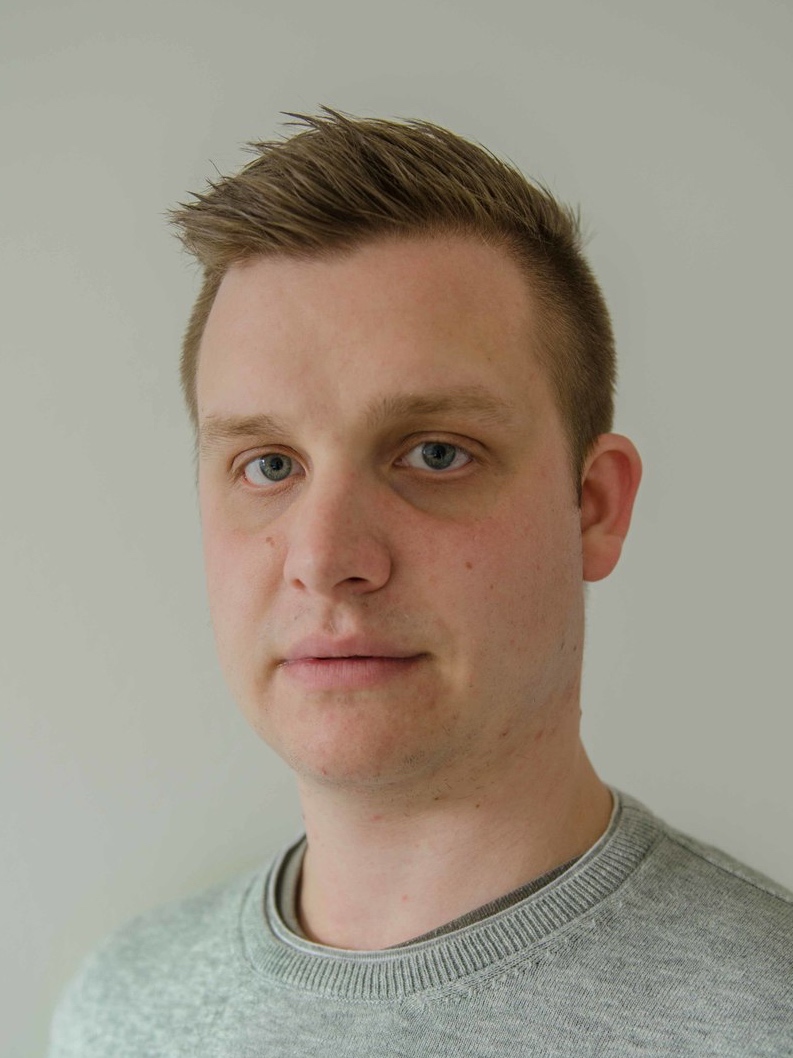}}]{Robert Schilling} received has B.Sc. and M.Sc. degrees in information and computer engineering from Graz University of Technology, where he is currently pursuing a Ph.D. degree. 
His current research interests include countermeasures against fault attacks, security enhanced processors and compilers, and software security.
\end{IEEEbiography}
% \newpage
\vspace{-2cm}
\begin{IEEEbiography}[{\includegraphics[width=0.85in,clip,keepaspectratio]{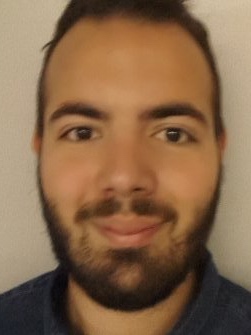}}]{Pasquale Davide Schiavone} received his B.Sc. (2013) and M.Sc. (2016) in computer engineering from Polytechnic of Turin. Since 2016 he has started his PhD studies at the Integrated Systems Laboratory, ETH Zurich. His research interests include low-power microprocessors design in multi-core systems and deep-learning architectures for energy-efficient systems.
\end{IEEEbiography}
% \newpage
\vspace{-2cm}
\begin{IEEEbiography}[{\includegraphics[width=0.85in,clip,keepaspectratio]{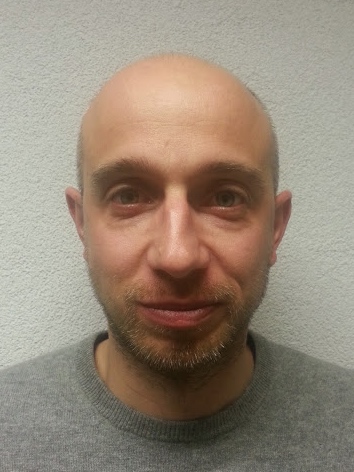}}]{Antonio Pullini}
received the M.S. degree in electrical engineering from Bologna University, Italy, and he is currently pursuing a Ph.D. degree from the Integrated Systems Laboratory, ETH Z\"urich, Switzerland. He was a Senior Engineer at iNoCs S.\`{a}.r.l., Lausanne, Switzerland. His research interests include low-power digital design and networks on chip.
\end{IEEEbiography}
% \newpage
\vspace{-2cm}
\begin{IEEEbiography}[{\includegraphics[width=0.85in,clip,keepaspectratio]{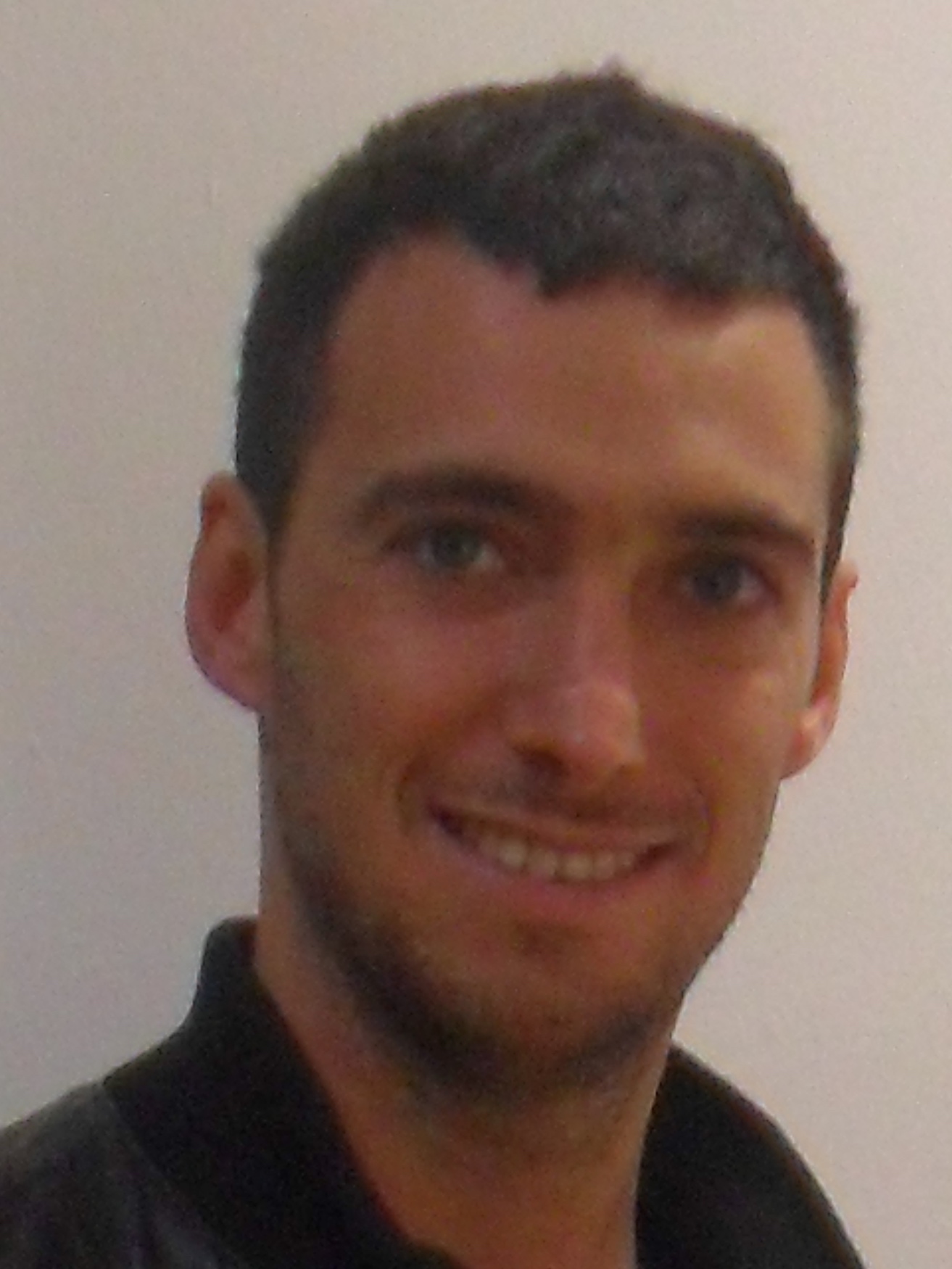}}]{Davide Rossi} is an assistant professor at the Energy Efficient Embedded Systems Laboratory at the University of Bologna. His current research interests include ultra-low-power multicore SoC design and applications. He has published more than 50 papers on international conferences and journals.
\end{IEEEbiography}
\newpage
% \vspace{-2cm}
\begin{IEEEbiography}[{\includegraphics[width=0.85in,clip,keepaspectratio]{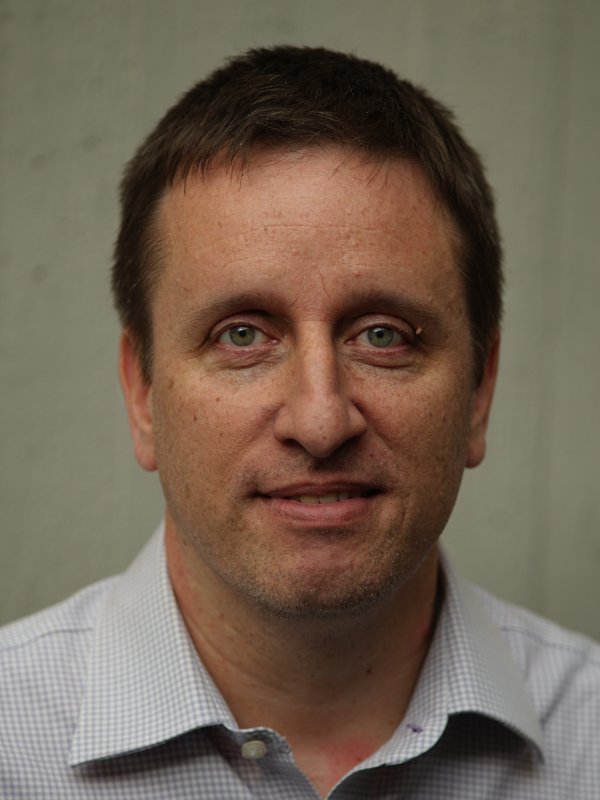}}]{Frank Ka\u{g}an G\"urkaynak} obtained his B.Sc. and M.Sc. degrees from Electrical 
and Electronical Engineering Department of the Istanbul Technical University and his Ph.D. degree from ETH Z\"urich. He is employed by the Microelectronics Design Center of ETH Zürich and his research interests include design of VLSI systems, cryptography, and energy efficient processing systems.
\end{IEEEbiography}
% \newpage
\vspace{-2cm}
\begin{IEEEbiography}[{\includegraphics[width=0.85in,clip,keepaspectratio]{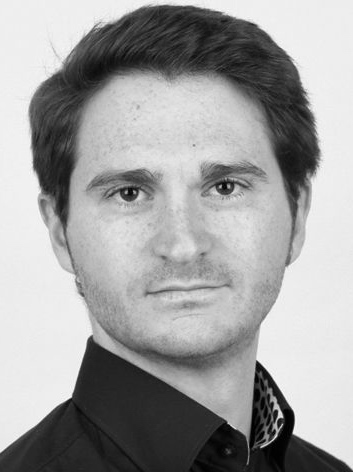}}]{Michael Muehlberghuber} received the MSc degree from Graz University of Technology in 2011 and the Ph.D. degree from ETH Zurich in 2017.
His research interests include hardware Trojans and the development of VLSI architectures of cryptographic primitives, targeting resource-constrained environments and high-performance applications.
\end{IEEEbiography}
% \newpage
\vspace{-2cm}
\begin{IEEEbiography}[{\includegraphics[width=0.85in,clip,keepaspectratio]{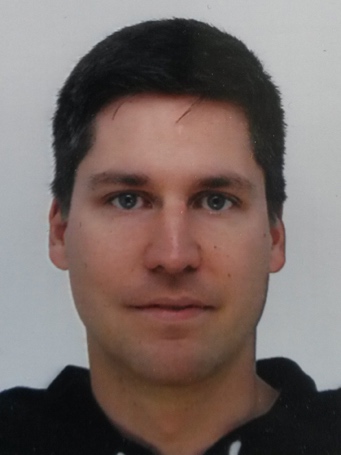}}]{Michael Gautschi}
received the M.Sc. degree in electrical engineering and information technology from ETH Z\"urich, Switzerland, in 2012.
Since then he has been with the Integrated Systems Laboratory, ETH Z\"urich, pursuing a Ph.D. degree.
His current research interests include energy-efficient systems, multi-core SoC design, mobile communication, and low-power integrated circuits.
\end{IEEEbiography}
% \newpage
\vspace{-2cm}
\begin{IEEEbiography}[{\includegraphics[width=0.85in,clip,keepaspectratio]{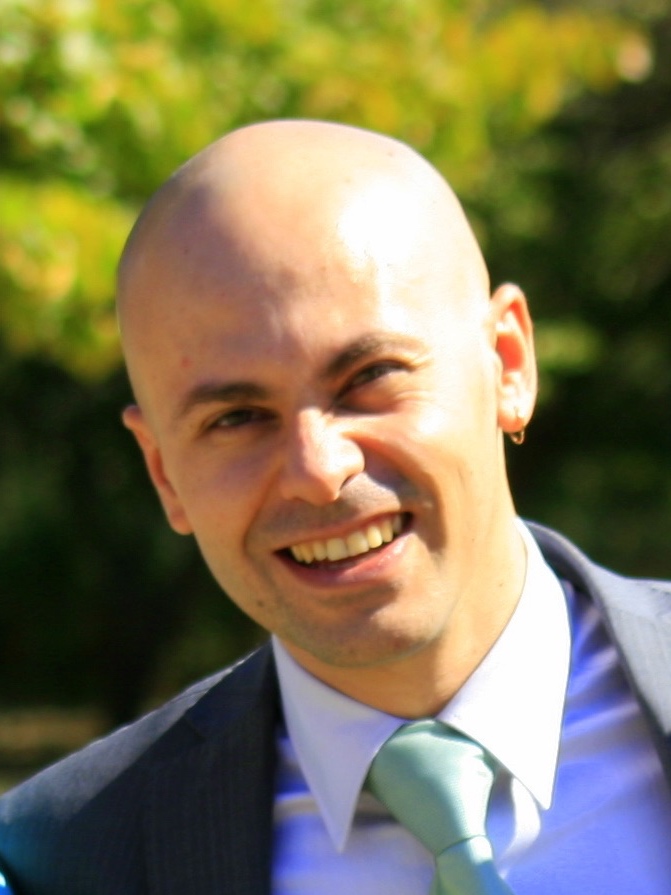}}]{Igor Loi} received the B.Sc. degree in electrical engineering from the University of Cagliari, Italy, in 2005, and the Ph.D. degree from the Department of Electronics and Computer Science, University of Bologna, Italy, in 2010.
He currently holds a researcher position in electronics engineering with the University of Bologna.
His research interests include ultra-low power multicore systems, memory system hierarchies, and ultra low-power on-chip interconnects.
\end{IEEEbiography}
% \newpage
\vspace{-2cm}
\begin{IEEEbiography}[{\includegraphics[width=0.85in,clip,keepaspectratio]{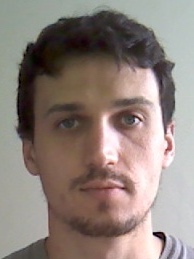}}]{Germain Haugou}
received the Engineering Degree in Telecommunication from the University of Grenoble (INP Ensimag), in 2004. He worked for 10 years in STMicroelectronics as a research engineer. He is currently working at the Swiss Federal Institute of Technology in Zurich, Switzerland, as a research assistant. His research interests include virtual platforms, run-time systems, compilers and programming models for many-core embedded architectures.
\end{IEEEbiography}
% \newpage
\vspace{-2cm}
\begin{IEEEbiography}[{\includegraphics[width=0.85in,clip,keepaspectratio]{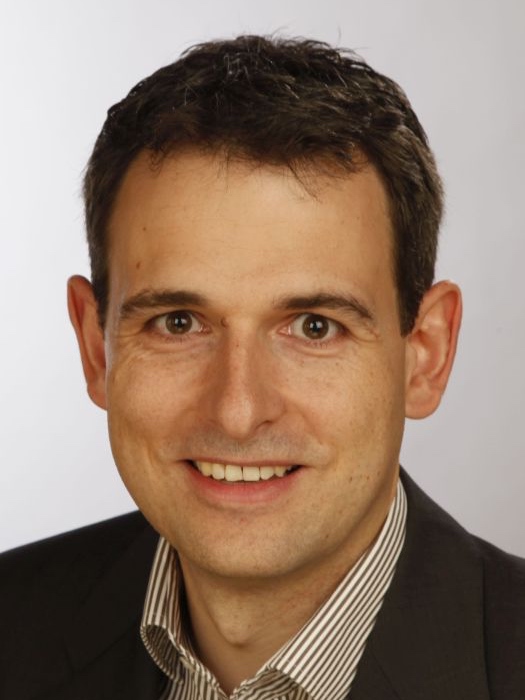}}]{Stefan Mangard} is a professor at TU Graz since 2013 and heads the
Secure Systems Group.
His research focuses on device and system security.
This in particular includes hardware/software security architectures, physical attacks and countermeasures,
microarchitectural attacks and countermeasures, cryptography, as well as secure and efficient
hardware and software implementations of cryptography.
\end{IEEEbiography}
% \newpage
\vspace{-2cm}
\begin{IEEEbiography}[{\includegraphics[width=0.85in,clip,keepaspectratio]{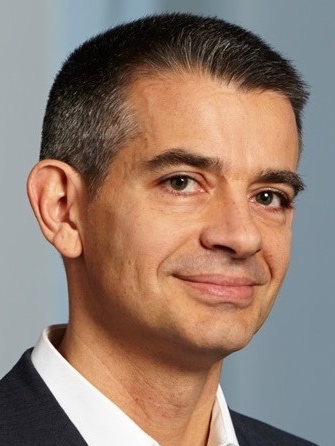}}]{Luca Benini} holds the chair of Digital Circuits and Systems at ETH Z\"{u}rich and is Full Professor at the Universit\`{a} di Bologna.
Dr. Benini's research interests are in energy-efficient system design for embedded and high-performance computing. 
He has published more than 800 papers, five books and several book chapters.
He is a Fellow of the ACM and a member of the Academia Europaea. He is the recipient of the 2016 IEEE CAS Mac Van Valkenburg award.
\end{IEEEbiography}

\end{document}